\documentclass[a4paper,11pt]{article}

\usepackage{amsmath,amssymb,amscd,bbm,amsfonts}
\usepackage{graphicx}
\usepackage{multirow}
\usepackage{cite}
\makeatletter

 \@addtoreset{equation}{section}
\makeatother

\newcounter{Enumerate}

\DeclareFontFamily{U}{rsf}{}
\DeclareFontShape{U}{rsf}{m}{n}{
  <5> <6> rsfs5 <7> <8> <9> rsfs7 <10-> rsfs10}{}
\DeclareMathAlphabet\Scr{U}{rsf}{m}{n}

\usepackage[mathscr]{eucal}

\newcommand{\vev}[1]{\langle {#1} \rangle}
\newcommand{\Vev}[1]{\big{\langle} {#1} \big{\rangle}}

\newcommand{\del}{\partial}
\newcommand{\half}{\frac{1}{2}}

\newcommand{\LS}{\ \ \ \ \ \ \ \ \ \ }
\newcommand{\ls}{\ \ \ \ \ }
\newcommand{\wt}{\widetilde}

\newcommand{\ve}{\varepsilon}
\newcommand{\ol}{\overline}

\newcommand{\dps}{\displaystyle}
\newcommand{\bsubeq}{\begin{subequations}}
\newcommand{\esubeq}{\end{subequations}}

\newcommand{\w}{\wedge}
\renewcommand{\d}{{\rm d}}
\newcommand{\nn}{\nonumber}
\newcommand{\e}{{\rm e}}
\newcommand{\slb}{\scalebox}
\newcommand{\Hf}{H^{\text{fl}}}
\newcommand{\dH}{\d_{\Hf}} 

\makeatletter
\def\bcline#1{\@bcline#1\@nil}
\def\@bcline#1-#2\@nil{%
  \omit
  \@multicnt#1%
  \advance\@multispan\m@ne
  \ifnum\@multicnt=\@ne\@firstofone{&\omit}\fi
  \@multicnt#2%
  \advance\@multicnt-#1%
  \advance\@multispan\@ne
  \cleaders\hbox{$\m@th \mbox{\rule{.1em}{.035em}}\mkern2mu$}\hfill
  \cr
  \noalign{\vskip-\arrayrulewidth}}
\def\multispan{\omit\@multispan}
\def\@multispan#1{%
  \@multicnt#1\relax
  \loop\ifnum\@multicnt>\@ne \sp@n\repeat}
\def\sp@n{\span\omit\advance\@multicnt\m@ne}
\makeatother

\begin{document}
\allowdisplaybreaks{

\thispagestyle{empty}


\begin{flushright}
arXiv:0810.0937 [hep-th] \\
YITP-08-91 \\
\end{flushright}

\vspace{20mm}

\begin{center}
\slb{1.75}{AdS Vacua, Attractor Mechanism}

\vspace{3mm}

\slb{1.75}{and Generalized Geometries}


\vspace{15mm}

\slb{1.2}{Tetsuji {\sc Kimura}}

\vspace{2mm}

\slb{1}{\sl
Yukawa Institute for Theoretical Physics, Kyoto University}

\slb{1}{\sl Sakyo-ku, Kyoto 606-8502, Japan}

\vspace{1mm}

\slb{0.9}{\tt tetsuji@yukawa.kyoto-u.ac.jp}

\end{center}

\vspace{18mm}


\begin{abstract}
We consider flux vacua attractor equations in type IIA string theory
compactified on generalized geometries with orientifold
projections. The four-dimensional ${\cal N}=1$ superpotential in this
compactification can be written as the sum of the Ramond-Ramond
superpotential and a term described by (non)geometric flux charges. We
exhibit a simple model in which supersymmetric AdS and Minkowski
solutions are classified by means of discriminants of the two
superpotentials. We further study various configurations without
Ramond-Ramond flux charges. In this case we find supersymmetric AdS
vacua both in the case of compactifications on generalized geometries
with $SU(3) \times SU(3)$ structures and on manifolds with an $SU(3)$-structure
without nongeometric flux charges. In the latter case, we have to
introduce correction terms into the prepotential in order to realize
consistent vacua. 
\end{abstract}

\newpage

\section{Introduction}

In the search of a model describing realistic physics, many
string compactification scenarios have been developed and studied.
One remarkable success is that of compactification on a Calabi-Yau
three-fold \cite{Candelas:1985en}, which gives a supersymmetric
Minkowski vacuum in four-dimensional effective gauge theory.
However, this configuration is insufficient as a candidate of a
realistic physical vacuum in string theory because it assumes four
significant simplifications:
constant dilaton, vanishing $H$-flux, flat Minkowski space, 
and ${\cal N}=1$ supersymmetry.
Once some of these assumptions are relaxed, a rich structure emerges in the
compactified space, which also affects the four-dimensional effective
theory. 
In particular, on a six-dimensional internal space with
$SU(3)$-structure \cite{Strominger:1986uh}, 
a non-vanishing NS-NS three-form flux $H$ yields torsion.
Such geometries have been investigated both in mathematics
\cite{Chiossi:2002} and in string theory \cite{Gauntlett:2003cy}.
Furthermore,
Hitchin's generalized geometry \cite{Hitchin:2002}
contains information about the $SU(3)$-structure manifold with torsion,
and provides a powerful technique 
in the investigation of four-dimensional ${\cal N}=2$ and ${\cal N}=1$ 
supergravity theories (see \cite{Grana:2005sn, Grana:2006hr,
  Koerber:2005qi, Martucci:2005ht, Grana:2005jc, Benmachiche:2006df,
  Grana:2006kf, Micu:2007rd, Koerber:2007xk,
  Cassani:2007pq, Grana:2008yw} and references therein).

Four-dimensional ${\cal N}=2$ (gauged) supergravity is not only dynamical 
but also controllable by two moduli spaces, namely
a special geometry and a quaternionic geometry \cite{Andrianopoli:1996cm}.
Generalized geometry also has two moduli spaces described as special geometries.
Due to the existence of these moduli spaces,
one can embed the four-dimensional ${\cal N}=2$ supergravity into type
IIA (or IIB) string theory
compactified on a generalized geometry. 
Various functions in four-dimensional spacetime such as the K\"{a}hler
potential and the superpotential are
written in terms of the prepotentials on the moduli spaces and
of various fluxes such as geometric fluxes and form fluxes
on the internal space.
The most generic forms of these functions are described by
Gra\~{n}a, Louis and Waldram 
\cite{Grana:2006hr},
and Benmachiche and Grimm demonstrated a consistent procedure to
truncate the model from ${\cal N}=2$ to ${\cal N}=1$ supersymmetry via 
an orientifold projection on generalized geometry \cite{Benmachiche:2006df}.
Gra\~{n}a, Minasian, Petrini and Tomasiello performed
a clever application of ``scanning'' technique to
${\cal N}=1$ vacua on parallelizable nilmanifolds
and solvmanifolds described as generalized geometries with
a single $SU(3)$-structure \cite{Grana:2006kf}.
In \cite{Cassani:2007pq} Cassani and Bilal carefully investigated 
the K\"{a}hler potential and superpotential in four-dimensional ${\cal N}=1$
supergravity obtained from type IIA string theory compactified
on generalized geometry with $SU(3) \times SU(3)$ structures.

String compactifications in the presence of fluxes also give rise to
non-abelian gauge symmetries in four-dimensional models, 
whereas the compactification on a Calabi-Yau space does not.
In order to realize such a gauge symmetry,
one introduces a twist in the (generalized) Scherk-Schwarz
compactification procedure \cite{Kaloper:1999yr, Dall'Agata:2005ff, Hull:2005hk},
possibly on some extended internal space, 
which yields
``nongeometric'' fluxes \cite{Kachru:2002sk,Shelton:2005cf} as well as
geometric fluxes.
One candidate for the internal space is 
generalized geometry with $SU(3) \times SU(3)$ structures
\cite{Grana:2006hr, Grana:2008yw}.
Other techniques, such as
compactification in the framework of doubled space formalism
\cite{Hull:2004in, Hull:2007jy}, have also been investigated to explain
nongeometric fluxes as arising from string dualities.

The aim of this paper is to realize 
consistent supersymmetric Anti-de Sitter (AdS) vacua
as well as Minkowski vacua
in type IIA theory with or without
Ramond-Ramond fluxes
\cite{Lust:2004ig, LopesCardoso:2002hd, LopesCardoso:2003sp, Acharya:2006ne, 
  Koerber:2008rx, Caviezel:2008ik}.
One way to obtain such vacua is by use of the attractor mechanism.
Originally the attractor mechanism was developed in the analysis of
the entropy of extremal (non-)BPS black holes in type II theories
\cite{Ferrara:1995ih, Ferrara:1996dd, Ferrara:1997tw,
Kallosh:2006bt, Hsu:2006vw, Ceresole07wx, Astefanesei:2007bf}.
This mechanism has been applied in the search of flux vacua 
\cite{Giddings:2001yu, Giryavets:2003vd,
  DeWolfe:2005uu, Kallosh:2005ax, Dall'Agata:2006nr, Bellucci:2007ds, Anguelova:2008fm},
since the functions and equations 
in both black holes and flux vacua frameworks are quite similar. 
In the black hole attractors one focuses on the black hole potential 
\cite{Ferrara:1997tw}, while
in the flux vacua attractors one studies the scalar potential
in ${\cal N}=1$ supergravity \cite{Giddings:2001yu}.
In both cases one investigates 
extreme points (called attractor points)
by evaluating the potentials, 
which are expressed in terms of the ${\cal N}=1$ superpotential.
The scalar potential is 
described in terms of the K\"{a}hler potential $K$ and
the superpotential ${\cal W}$ as
\begin{align}
V \ &= \ 
\e^K \big( K^{M \ol{N}} D_M {\cal W} D_{\ol{N}} \ol{\cal W} 
- 3 |{\cal W}|^2 \big)
+ \half ({\rm Re} f)_{\hat{a} \hat{b}} D^{\hat{a}} \ol{D}{}^{\hat{b}}
\ \equiv \ 
V_{\cal W} + V_D
\, , \label{SV}
\end{align}
where $D_M$ is the K\"{a}hler covariant derivative
with respect to complex scalar fields
$\phi^M$, defined as $D_M {\cal W} \equiv (\del_M + \del_M K) {\cal W}$,
and $K_{M \ol{N}} = \del_M \del_{\ol{N}} K (\phi, \ol{\phi})$ is the
K\"{a}hler metric.
The $\phi^M$ collectively denote all complex scalars in all
chiral multiplets present in the ${\cal N}=1$ theory. 
The second term on the right-hand side 
carries the D-terms $D^{\hat{a}}$ which belong to vector multiplets.
The attractor point is defined by the equation
$\phi^M = \phi_*^M$ satisfying $\del V/ \del \phi^M |_{*} = 0$.

This paper is structured as follows:
In section \ref{section-Basic}
we write down the scalar potential and its derivatives 
in four-dimensional ${\cal N}=1$ supergravity.
We evaluate the derivatives of the scalar potential, 
which are called attractor equations.
In order to make our discussion clear, 
we restrict the prepotential 
governing
the superpotential to a simple form.
In sections \ref{section-generic-model}, 
\ref{section-RR0}, \ref{section-RR0-nongeom0} and \ref{section-RR0-nongeom0-deform},
we find flux vacua attractors in various examples.
In section \ref{section-generic-model} we analyze a model in
which Ramond-Ramond fluxes as well as (non)geometric
fluxes are introduced. 
In this analysis two discriminants of the superpotential 
play central roles in the classification of supersymmetric vacua,
where the discriminants are written in terms of flux charges.
If the discriminants are positive, 
we obtain a supersymmetric AdS vacuum whose cosmological
constant is governed by (the square root of) a discriminant of the superpotential. 
If the discriminants are negative,
we obtain a supersymmetric Minkowski vacuum. 
In section \ref{section-RR0}
we analyze a different model in which Ramond-Ramond flux charges are
absent, whereas nongeometric flux charges are present.
There we again obtain a supersymmetric AdS vacuum.
In section \ref{section-RR0-nongeom0}
we study other models which carry only geometric flux charges,
where we find neither supersymmetric nor non-supersymmetric solutions if the
prepotential is expressed only in terms of the intersection number, 
as in the case of Calabi-Yau compactification in the large volume limit.
In section \ref{section-RR0-nongeom0-deform}
we introduce correction terms to the prepotential 
in order to find supersymmetric 
vacua in the presence of geometric flux charges, but without
nongeometric and Ramond-Ramond flux charges.
We can interpret each of the models in this section 
as coming from heterotic string
theory compactified on a torsionful manifold with a single $SU(3)$-structure.
Section \ref{Discussions} is devoted to the summary and discussions.
To streamline the arguments, brief derivations of the functions in section
\ref{section-Basic} are included in appendices.


\section{Analysis of scalar potential}
\label{section-Basic}



In this section 
we analyze the scalar potential (\ref{SV}).
We start with type IIA string theory compactified on generalized
geometry with $SU(3) \times SU(3)$ structures.
This compactification yields the superpotential ${\cal W}$, the
K\"{a}hler potential $K$, the dilaton $\varphi$ and the D-terms
$D^{\hat{A}}$ in four-dimensional spacetime.
Their explicit forms are 
\bsubeq \label{IIAKWN=1}
\begin{align}
{\cal W} \ &= \ 
- \frac{i}{4 \ol{a}b} \Big[
X^{\check{A}} \big( 
e_{\text{RR} \check{A}} - U^{\check{I}} e_{\check{I} \check{A}} 
+ \wt{U}_{\hat{I}} m_{\check{A}}{}^{\hat{I}} 
\big)
- {\cal F}_{\check{A}} \big( 
m_{\text{RR}}^{\check{A}} + U^{\check{I}} p_{\check{I}}{}^{\check{A}} 
- \wt{U}_{\hat{I}} q^{\hat{I} \check{A}} 
\big)
\Big]
\, , \label{IIASuperN=1} \\
K \ &= \ 
K_+ + 4 \varphi
\, , \label{IIAKahlerN=1} \\
K_+ \ &= \ 
- \log i \big( 
\ol{X}{}^{\check{A}} {\cal F}_{\check{A}} - X^{\check{A}} \ol{\cal F}_{\check{A}} 
\big) 
\, , \\
\e^{- 2 \varphi} \ &= \ 
\half \Big[ {\rm Im} ({\cal C} Z^{\check{I}}) {\rm Re} ({\cal C} {\cal G}_{\check{I}})
- {\rm Re} ({\cal C} Z^{\hat{I}}) {\rm Im} ({\cal C} {\cal G}_{\hat{I}}) \Big]
\, , \label{IIAdilatonN=1} \\
D^{\hat{A}}
\ &= \ 
\e^{2 \varphi} [ ({\rm Im} {\cal N})^{-1} ]^{\hat{A} \hat{B}} 
\Big\{ {\rm Re} ({\cal C} Z^{\hat{I}}) 
\big[ e_{\hat{I}\hat{B}} + {\cal N}_{\hat{B}\hat{C}} p_{\hat{I}}{}^{\hat{C}} \big]
- {\rm Re} ({\cal C} {\cal G}_{\check{I}} ) 
\big[ m_{\hat{B}}{}^{\check{I}} + {\cal N}_{\hat{B}\hat{C}} q^{\check{I}\hat{C}} \big]
\Big\}
\, . \label{IIAD1}
\end{align}
\esubeq
Here we used notation and conventions in \cite{Grimm:2004ua,
  Grana:2006kf, Grana:2006hr, Cassani:2007pq}.
We summarize derivations of the above functions in the appendices. 



Let us search an extreme point of the scalar potential given by 
$\del_P V = 0$ with respect to holomorphic variables.
The first derivatives are written as
\bsubeq \label{1deriv-V}
\begin{align}
\del_P V_{\cal W}
\ &= \ 
\e^K \Big\{
K^{M \ol{N}} D_P D_M {\cal W} D_{\ol{N}} \ol{\cal W}
+ \del_P K^{M \ol{N}} D_M {\cal W} D_{\ol{N}} \ol{\cal W} 
- 2 \ol{\cal W} D_P {\cal W}
\Big\}
\, , \label{1deriv-VW} \\
\del_P V_D \ &= \
\half \del_P ({\rm Re} f)_{\hat{a} \hat{b}} D^{\hat{a}} \ol{D}{}^{\hat{b}}
+ \half ({\rm Re} f)_{\hat{a} \hat{b}} \del_P D^{\hat{a}} \ol{D}{}^{\hat{b}}
+ \half ({\rm Re} f)_{\hat{a} \hat{b}} D^{\hat{a}} \del_P \ol{D}{}^{\hat{b}}
\, , \label{1deriv-VD}
\end{align} 
\esubeq
where we used $\del_P \ol{\cal W} = 0$ and a set of equations:
\bsubeq
\begin{align}
\ol{\cal W} \del_P {\cal W}
\ &= \ 
\ol{\cal W} D_P {\cal W} - \del_P K | {\cal W} |^2
\, , \\
D_P D_M {\cal W} 
\ &= \ 
\del_P D_M {\cal W} 
+ \del_P K D_M {\cal W} 
\, , \\
\begin{split}
D_P D_{\ol{N}} \ol{\cal W}
\ &= \ 
K_{P \ol{N}} \ol{\cal W}
\, .
\end{split}
\end{align}
\esubeq
The K\"{a}hler covariant derivative is defined 
in terms of the K\"{a}hler potential 
$K = K_+ + 4 \varphi$.
This does {\it not} inherit the property of
the special K\"{a}hler geometry of local type. 

We look for a solution which satisfies $\del_P V_{\cal W} = 0$
and $\del_P V_D = 0$.
This is realized when the supersymmetry condition $D_P {\cal W} = 0$ is satisfied.
The equation $D_P {\cal W} = 0$ is called the attractor equation
in supersymmetric attractor mechanism.
The holomorphic scalar fields are described by 
\bsubeq
\begin{align}
t^{\check{a}} \ &= \ 
\frac{X^{\check{a}}}{X^0} \ = \ 
b^{\check{a}} + i v^{\check{a}} 
\, , \ls
U^{\check{I}} \ = \ \xi^{\check{I}} + i \, {\rm Im} ({\cal C} Z^{\check{I}})
\, , \ls
\wt{U}_{\hat{I}} \ = \ 
\wt{\xi}_{\hat{I}} + i \, {\rm Im} ({\cal C} {\cal G}_{\hat{I}})
\, , \\
I \ &= \ 0,1,\dots, b^-
\, \, \ls
\check{I} \ = \ 0,1,\dots, \check{b}{}^-
\, , \ls
\hat{I} \ = \ 1,\dots, \hat{b}{}^- 
\, , \ls 
\hat{b}{}^- \ \equiv \ b^- - \check{b}{}^- 
\, .
\end{align} 
\esubeq
Since $Z^0$ is compensated by the four-dimensional dilaton $\varphi$
via the combination ${\cal C} Z^0$ \cite{Cassani:2007pq},
$U^0 = \xi^0 + i {\rm Im} ({\cal C} Z^0)$ is dynamical.

In order to extract significant property of vacua,
it is much instructive to restrict the prepotentials.
Precisely speaking, 
we reduce the prepotential ${\cal F}$ on ${\cal M}_+$
and the number of degrees of freedom in the moduli space ${\cal M}_-$. 
Here we set the prepotential ${\cal F}$ 
\cite{Grimm:2004ua, Cassani:2007pq} 
in the following form: 
\begin{align}
{\cal F} \ &\equiv \ 
D_{abc} \frac{X^a X^b X^c}{X^0}
\, , \ls
D_{abc}
\ = \ 
- \frac{1}{6} {\cal K}_{abc} 
\, . \label{IIA-F3}
\end{align}
We should keep in mind that 
the expression (\ref{IIA-F3}) implies that all $\alpha'$ corrections are neglected.
In Calabi-Yau compactification this setting is usual, while 
quite restricted in the case of compactifications on generalized geometries.
We will discuss this issue in sections
\ref{section-RR0-nongeom0} and \ref{section-RR0-nongeom0-deform}. 
We also restrict another moduli space ${\cal M}_-$.
For simplicity, we reduce the number of moduli.
We set $\check{I} = \{ 0 \}$.
This means that the remaining dynamical field is $U^0$,
and we truncate out all of
$U^{\check{i}}$ and $\wt{U}_{\hat{I}}$.
From now on we abbreviate $U^0$ to $U$.
As far as we concern this reduction, the constraints (\ref{nongeom-Q-charges})
are trivial.


\subsection{Derivatives of superpotential}

The superpotential (\ref{IIASuperN=1}) 
is governed by the Ramond-Ramond flux charges and the (non)geometric flux charges.
Its explicit form is
\bsubeq \label{IIA-WRR-WQ}
\begin{gather}
{\cal W}
\ = \ 
{\cal W}^{\text{RR}} + U {\cal W}^{\cal Q}
\, , \\
{\cal W}^{\text{RR}}
\ \equiv \ 
- \frac{i}{4 \ol{a}b} \Big(
X^{\check{A}} e_{\text{RR} \check{A}} 
- {\cal F}_{\check{A}} \, m_{\text{RR}}^{\check{A}} 
\Big)
\, , \ls
{\cal W}^{\cal Q}
\ \equiv \ 
\frac{i}{4 \ol{a}b} 
\Big(
X^{\check{A}} e_{0 \check{A}} 
+ {\cal F}_{\check{A}} \, p_0{}^{\check{A}} 
\Big)
\, .
\end{gather}
\esubeq
We refer to ${\cal W}^{\text{RR}}$ as the Ramond-Ramond flux superpotential,
and to ${\cal W}^{\text{RR}}$ as the (non)geometric flux superpotential. 
The K\"{a}hler covariant derivatives acting on the superpotential are
\bsubeq \label{IIA-1cov-W}
\begin{align}
D_{\check{a}} {\cal W}
\ &= \
D_{\check{a}} {\cal W}^{\text{RR}}
+ U D_{\check{a}} {\cal W}^{\cal Q}
\, , \label{IIA-DaW} \\
D_{U} {\cal W}
\ &= \ 
\frac{i}{{\rm Im} U} \Big( {\cal W}^{\text{RR}} + {\rm Re} U {\cal W}^{\cal Q} \Big)
\, . \label{IIA-DUW}
\end{align}
\esubeq
We study the second K\"{a}hler covariant derivatives of the
superpotential $D_M D_N {\cal W} =  
\del_M D_N {\cal W} + \del_M K D_N {\cal W}$.
The explicit forms are 
\bsubeq \label{IIA-DDW}
\begin{align}
D_{\check{b}} D_{\check{a}} {\cal W}
\ &= \
D_{\check{b}} D_{\check{a}} {\cal W}^{\text{RR}}
+ U D_{\check{b}} D_{\check{a}} {\cal W}^{\cal Q}
\ = \
i C_{\check{a} \check{b} \check{c}} (K_+)^{\check{c} \ol{\check{d}}}
\Big( D_{\ol{\check{d}}} \ol{\cal W}{}^{\text{RR}}
+ U D_{\ol{\check{d}}} \ol{\cal W}{}^{\cal Q}
\Big)
\, , \label{IIA-DaDbW} \\
D_{U} D_{\check{a}} {\cal W}
\ &= \
D_{\check{a}} {\cal W}^{\cal Q}
+ \frac{i}{{\rm Im} U}
D_{\check{a}} {\cal W}
\, , \label{IIA-DUDaW} \\
D_{U} D_{U} {\cal W}
\ &= \ 
\frac{i}{2 {\rm Im} U}
\Big( 3 D_U {\cal W} + {\cal W}^{\cal Q} \Big)
\, . \label{IIA-DUDUW}
\end{align}
\esubeq
Since $t^a$ is independent of $U$, 
we can use a formula 
$D_a D_b \big( \e^{\frac{K_+}{2}} {\cal W} \big) = 
i C_{abc} (K_+)^{c \ol{d}} D_{\ol{d}} \big( \e^{\frac{K_+}{2}} \ol{\cal W} \big)$. 
As far as we concern
the system with the prepotential (\ref{IIA-F3}), 
the imaginary part of $U$ does not vanish, otherwise 
the K\"{a}hler metric $K_{U \ol{U}}$ and the curvature
tensor $R^U{}_{UU \ol{U}}$ become singular.

Now we are ready to evaluate the extreme point of $V_{\cal W}$.
Due to the equations (\ref{IIA-1cov-W}), we obtain a set of differential
equations and an algebraic equation
at the extreme point $( t^{\check{a}} , U ) = ( t^{\check{a}}_*, U_* )$:
\bsubeq \label{IIA-AP-DW}
\begin{alignat}{3}
D_{\check{a}} {\cal W} \big|_* \ &= \ 0 
\ls &\to & \ls &
D_{\check{a}} {\cal W}^{\text{RR}} \big|_*
\ &= \ 
- U_* D_{\check{a}} {\cal W}^{\cal Q} \big|_*
\, , \ls \label{IIA-AP-DaW}  \\
D_U {\cal W} \big|_* \ &= \ 0 
\ls &\to & \ls &
{\cal W}_*^{\text{RR}} \ &= \ 
- {\rm Re} U_*
{\cal W}_*^{\cal Q}
\, .  \label{IIA-AP-DUW}
\end{alignat}
\esubeq
If both $D_{\check{a}} {\cal W}^{\text{RR}}$
and $D_{\check{a}} {\cal W}^{\cal Q}$ vanish to satisfy
(\ref{IIA-AP-DaW}) while ${\cal W}^{\text{RR}}$ and ${\cal W}^{\cal Q}$ do not vanish,
we obtain flux vacua attractor equations. 
These are exactly the same equations in the black hole attractors in type IIA
theory \cite{Ferrara:1996dd, Ferrara:1997tw, Kallosh:2006bt}.
On the other hand, if we can take $D_{\check{a}} {\cal W}^{\text{RR}} |_* \neq 0$
with $D_{\check{a}} {\cal W} |_* = 0$, 
we can employ the
non-supersymmetric black hole analyses in finding supersymmetric flux vacua.
The discussions of a classification of (non)supersymmetric vacua 
can be seen in \cite{Bellucci:2007ds}.


\subsection{Derivatives of D-term}

Investigation of the D-term  (\ref{IIAD1}) is interesting,
because its non-trivial value breaks supersymmetry.
It is known that 
\begin{align}
f_{\hat{a} \hat{b}} \ = \ 
- i \ol{\cal N}_{\hat{a} \hat{b}} 
\ = \ 
i {\cal K}_{\hat{a} \hat{b} \check{c}} t^{\check{c}}
\, , \ \ \ 
({\rm Re} f)_{\hat{a} \hat{b}} \ = \ 
- ({\rm Im} {\cal N})_{\hat{a} \hat{b}} \ = \ 
- {\cal K}_{\hat{a} \hat{b}} 
\, , \ \ \ 
({\rm Re} {\cal N})_{\hat{a} \hat{b}} \ = \ 
- {\cal K}_{\hat{a} \hat{b} \check{c}} b^{\check{c}}
\, . \label{IIA_N-f}
\end{align}
Here let us write a concrete form 
\begin{gather}
D^{\hat{a}}
\ = \ 
- \frac{2}{{\rm Im} U}
[ ({\rm Im} {\cal N})^{-1} ]^{\hat{a} \hat{b}} 
\Big( m_{\hat{b}}{}^0 + {\cal N}_{\hat{b} \hat{c}} \, q^{0\hat{c}} \Big)
\, .
\end{gather}
Note that the indices $\hat{A}$ are reduced\ to $\hat{a}$, since
the graviphoton $A_{\mu}^0$ is always truncated out. 
The intersection number 
${\cal K}_{\hat{a} \hat{b} \check{c}} = \int_{\mathscr{M}}
\omega_{\hat{a}} \w \omega_{\hat{b}} \w \omega_{\check{c}}$ is
a constant.
It is useful to introduce an inverse ${\cal K}^{\hat{a} \hat{b}}$
which satisfies a relation
${\cal K}^{\hat{a} \hat{b}} {\cal K}_{\hat{b} \hat{c}} = \delta^{\hat{a}}_{\hat{c}}$.
The D-term and the potential can be rewritten as
\begin{align}
\mu_{\hat{a}} 
\ &\equiv \
- ({\rm Im} {\cal N})_{\hat{a} \hat{b}} D^{\hat{b}}
\ = \ 
\frac{2}{{\rm Im} U}
\Big( m_{\hat{a}}{}^0 
- {\cal K}_{\hat{a} \hat{b} \check{c}} \ol{t}{}^{\check{c}} \, q^{0\hat{b}} \Big)
\, , \ls
V_D \ = \ 
- \half {\cal K}^{\hat{a} \hat{b}} \mu_{\hat{a}} \ol{\mu}_{\hat{b}}
\, .
\end{align}
The first derivative $\del_P V_D$ is written as
$\del_P V_D
= 
- \half \del_P {\cal K}^{\hat{a} \hat{b}} \mu_{\hat{a}} \ol{\mu}_{\hat{b}}
- \half {\cal K}^{\hat{a} \hat{b}} \del_P \mu_{\hat{a}} \ol{\mu}_{\hat{b}}
- \half {\cal K}^{\hat{a} \hat{b}} \mu_{\hat{a}} \del_P \ol{\mu}_{\hat{b}}$.
The derivatives depend only on the complex variables 
$t^{\check{a}} = b^{\check{a}} + i v^{\check{a}}$:
\begin{align}
\del_{\check{c}} {\cal K}_{\hat{a} \hat{b}}
\ &= \ 
{\cal K}_{\hat{a} \hat{b} \check{d}} \frac{\del v^{\check{d}}}{\del t^{\check{c}}}
\ = \ 
- i {\cal K}_{\hat{a} \hat{b} \check{c}}
\, , &
\del_{\check{c}} {\cal N}_{\hat{a} \hat{b}}
\ &= \ 
- {\cal K}_{\hat{a} \hat{b} \check{d}} \frac{\del \ol{t}{}^{\check{d}}}{\del t^{\check{c}}}
\ = \ 
0 , &
\del_{\check{c}} \ol{\cal N}_{\hat{a} \hat{b}}
\ &= \ 
- {\cal K}_{\hat{a} \hat{b} \check{c}}
\, . \label{IIA-1deriv-Ref} 
\end{align}
We also study the first derivatives of $\mu_{\hat{a}}$:
\begin{align}
\del_{\check{b}} \mu_{\hat{a}}
\ &= \ 
0
\, , & 
\del_{\check{b}} \ol{\mu}_{\hat{a}}
\ &= \ 
- \frac{2}{{\rm Im} U}
{\cal K}_{\hat{a} \hat{c} \check{b}} q^{0 \hat{c}}
\, , & 
\del_{U} \mu_{\hat{a}}
\ &= \
\frac{i}{2 {\rm Im} U}
\mu_{\hat{a}}
\, , & 
\del_{U} \ol{\mu}_{\hat{a}}
\ &= \
\frac{i}{2 {\rm Im} U}
\ol{\mu}_{\hat{a}}
\, . \label{IIA-1deriv-M}
\end{align}
Then we obtain
\begin{align}
\del_{\check{c}} V_D 
\ &= \
- \frac{i}{2} {\cal K}^{\hat{a} \hat{e}} {\cal K}_{\check{c} \hat{d} \hat{e}}
\, \mu_{\hat{a}} \Big( 
{\cal K}^{\hat{b} \hat{d}} \ol{\mu}_{\hat{b}} 
+ \frac{2i}{{\rm Im} U}
q^{0 \hat{d}} \Big)
\; , \ls
\del_{U} V_D 
\ = \ 
- \frac{i}{4 {\rm Im} U}
{\cal K}^{\hat{a} \hat{b}} 
\, \mu_{\hat{a}} \ol{\mu}_{\hat{b}}
\, .
\end{align}
Since there are no contributions of $m_{\hat{a}}{}^0$ and $q^{0 \hat{a}}$ 
to the scalar potential $V_{\cal W}$, we can evaluate the extreme
point of $V_D$ independently.
If we consider the condition $\del_P V_D |_* = 0$, we find that
\begin{align}
\mu_{\hat{a}} \big|_* \ &= \ 0
\, \label{IIA-AP-D}
\end{align}
is the solution. 
This implies that the D-term contribution does not break supersymmetry
and the scalar potential $V_D$ vanishes at the extreme point.
Then it is enough to focus on the scalar potential 
$V_{\cal W}$ in order to analyze whether supersymmetry of the
effective theory is broken or not.


\subsection{Search of (non)supersymmetric flux attractor vacua}
\label{comments-section}

So far we specified the K\"{a}hler potential, the superpotential,
the D-term and their derivatives. 
In next sections we study various simple examples which show intrinsic phenomena
in supersymmetric flux vacua. 
First, we consider a setup which contains 
not only Ramond-Ramond fluxes but
also (non)geometric fluxes.
In this setup we obtain a simple but powerful rule to find
supersymmetric vacua.
Second, we study other cases in which the Ramond-Ramond flux charges
are absent. More precisely we consider the following three cases:
(i) No Ramond-Ramond flux charges in the presence of the nongeometric flux
charges:
(ii) No Ramond-Ramond flux charges in the absence of the nongeometric flux
charges without any corrections in the prepotential (\ref{IIA-F3}):
(iii) No Ramond-Ramond flux charges in the absence of the nongeometric flux
charges with corrections in the prepotential.
Indeed a generalized geometry with neither the Ramond-Ramond fluxes nor 
the nongeometric fluxes corresponds to an $SU(3)$-structure manifold in
string flux compactification.
We do not consider other situations that 
all the (non)geometric flux charges vanish while there exist 
non-zero Ramond-Ramond flux charges.
These configurations are forbidden \cite{Lust:2004ig}
because the Ramond-Ramond fluxes induce the non-zero valued NS-NS flux
and some torsion classes.


\section{Example 1: a model with  Ramond-Ramond flux charges} 
\label{section-generic-model}

\subsection{Strategy}

In section \ref{section-Basic} we discussed
the attractor equations $D_P {\cal W} = 0$. 
Here we set $a = \ol{b} \, \e^{i \theta}$ and $|a|^2 = |b|^2 = \half$
as in (\ref{O6condition}) via the O6 orientifold projection.
An arbitrary parameter $\theta$ is absorbed in the phase of $a$ (or $b$) 
to set $2 \ol{a} b = -i$.
We rescale all the flux charges by integer $2$
without loss of generality.
The scalar potential on the extreme point is given by
\bsubeq
\begin{gather}
D_{\check{a}} {\cal W} \big|_* \ = \ 0
\; , \ls
D_U {\cal W} \big|_* \ = \ 0
\, , \label{IIA-AP-BBW} \\
V_* \ = \ 
\e^{K} \Big( K^{M \ol{N}} 
D_M {\cal W} D_{\ol{N}} \ol{\cal W}
- 3 |{\cal W}|^2 \Big)_*
- \half {\cal K}^{\hat{a} \hat{b}} \mu_{\hat{a}} \ol{\mu}_{\hat{b}} \Big|_*
\ = \ 
- 3 \, \e^{K} |{\cal W}_*|^2
\, .
\end{gather}
\esubeq
This gives a non-positive cosmological constant. 
The four-dimensional spacetime becomes 
a Minkowski space (if ${\cal W}_* = 0$) 
or an AdS space (if ${\cal W}_* \neq 0$).
Here let us consider a model governed by a single modulus
$t^{\check{a}} \equiv t$.
Various functions are simplified:
\begin{align}
K_+ \ &= \ - \log \big( - i D (t - \ol{t})^3 \big)
\, , \ls
C_{ttt} \ = \ 
\frac{6i}{(t - \ol{t})^3}
\, , \label{IIA-single-fns}
\end{align}
where we set $D_{\check{a} \check{b} \check{c}} = D$.
The superpotentials ${\cal W}^{\text{RR}}$ and 
${\cal W}^{\cal Q}$ (\ref{IIA-WRR-WQ}) are explicitly given by
\bsubeq \label{IIA-single-WRR-WQ}
\begin{align}
{\cal W}^{\text{RR}}
\ &= \ 
X^{\check{A}} e_{\text{RR} \check{A}} - {\cal F}_{\check{A}} m_{\text{RR}}^{\check{A}}
\ = \
e_{\text{RR}0} + e_{\text{RR}} t - 3 m_{\text{RR}} t^2 + m_{\text{RR}}^0 t^3
\, , \label{IIA-single-WRR} \\
{\cal W}^{\cal Q}
\ &= \ 
- \Big(
X^{\check{A}} e_{0 \check{A}} + {\cal F}_{\check{A}} p_0{}^{\check{A}}
\Big)
\ = \ 
- e_{00} - e_{0} t - 3 p_{0} t^2 + p_0{}^0 t^3
\, . \label{IIA-single-WQ}
\end{align}
\esubeq
For simplicity, we assume that $m_{\text{RR}}^0$ and $p_0{}^0$ are
positive definite and $D = 1$. 
To restore explicit contributions of $D$,
one replaces the charges $( m_{\text{RR}} , m_{\text{RR}}^0 , p_0 , p_0{}^0 )$ 
to $( D m_{\text{RR}} , D m_{\text{RR}}^0 , D p_0 , D p_0{}^0 )$.

Following the discussion in (\ref{IIA-AP-DW}), the supersymmetry
condition is described by one differential and one algebraic 
equations with respect to 
${\cal W}^{\text{RR}}$ and ${\cal W}^{\cal Q}$:
\bsubeq \label{IIA-AP-BBW2}
\begin{alignat}{3}
D_t {\cal W} \big|_* \ &= \ 0 
\ls &\to \ls &&
D_t {\cal W}^{\text{RR}} \big|_* \ &= \ 
- U_* D_t {\cal W}^{\cal Q} \big|_*
\, , \\
D_U {\cal W} \big|_* \ &= \ 0
\ls &\to \ls &&
{\cal W}_*^{\text{RR}} 
\ &= \ 
- {\rm Re} U_* {\cal W}_*^{\cal Q}
\, .
\end{alignat}
\esubeq
It is useful to consider discriminants\footnote{The author would
  like to thank Tohru Eguchi for his introducing an essential idea of
the usage of discriminants.} of the Ramond-Ramond flux superpotential
${\cal W}^{\text{RR}}$ and of its derivative $\del_t {\cal W}^{\text{RR}}$:
\bsubeq \label{IIA-WRR-discrim}
\begin{align}
\Delta ({\cal W}^{\text{RR}})
\ \equiv \ 
\Delta^{\text{RR}}
\ &= \ 
-27 \big( m_{\text{RR}}^0 e_{\text{RR}0} \big)^2 
- 54 m_{\text{RR}}^0 e_{\text{RR}0} m_{\text{RR}} e_{\text{RR}} 
+ 9 \big( m_{\text{RR}} e_{\text{RR}} \big)^2
\nn \\
\ & \ \ \ \
+ 108 (m_{\text{RR}})^3 e_{\text{RR}0} 
- 4 m_{\text{RR}}^0 (e_{\text{RR}})^3 
\, , \\
\Delta (\del_t {\cal W}^{\text{RR}})
\ \equiv \ 
\lambda^{\text{RR}}
\ &= \ 
12 \big( 3 (m_{\text{RR}})^2 - m_{\text{RR}}^0 e_{\text{RR}} \big)
\, .
\end{align}
\esubeq
It is also useful to discuss discriminants of 
the (non)geometric flux superpotential ${\cal W}^{\cal Q}$ and of its derivative 
$\del_t {\cal W}^{\cal Q}$:
\bsubeq \label{IIA-WQ-discrim}
\begin{align}
\Delta ({\cal W}^{\cal Q})
\ \equiv \ 
\Delta^{\cal Q}
\ &= \ 
-27 \big( p_0{}^0 e_{00} \big)^2 
- 54 p_0{}^0 e_{00} p_0 e_{0} 
+ 9 \big( p_0 e_0 \big)^2
- 108 (p_0)^3 e_{00}
+ 4 p_0{}^0 (e_0)^3 
\, , \\
\Delta (\del_t {\cal W}^{\cal Q})
\ \equiv \ 
\lambda^{\cal Q}
\ &= \ 
12 \big( 3 (p_0)^2 + p_0{}^0 e_0 \big)
\, .
\end{align}
\esubeq

Our strategy is as follows:
First we investigate zeros of the Ramond-Ramond flux
superpotential ${\cal W}^{\text{RR}}$ and those of its covariant derivative
$D_t {\cal W}^{\text{RR}}$ 
by using the discriminants $\Delta^{\text{RR}}$ and $\lambda^{\text{RR}}$.
Second we analyze the (non)geometric flux superpotential
${\cal W}^{\cal Q}$ in terms of the discriminants
$\Delta^{\cal Q}$ and $\lambda^{\cal Q}$ in a parallel way.
Third we evaluate possible supersymmetric vacua following the
equations (\ref{IIA-AP-BBW2}).


\subsection{Ramond-Ramond flux superpotential}
\label{IIA-single-WR-generic}

\subsubsection{Solutions of $D_t {\cal W}^{\rm RR} = 0$}

We formally describe a solution of 
$D_t {\cal W}^{\text{RR}} = 0$:
\bsubeq \label{IIA-AP-DWR0}
\begin{align}
t_* \ &\equiv \ t_{1*} + i t_{2*}
\ = \ 
\frac{6(3 m_{\text{RR}}^0 e_{\text{RR}0} + m_{\text{RR}} e_{\text{RR}})}{\lambda^{\text{RR}}}
\pm \frac{2 i \sqrt{3 \Delta^{\text{RR}}}}{\lambda^{\text{RR}}}
\, . \label{IIA-AP-WR-t_sol}
\end{align}
The superpotential at this point is given by
\begin{align}
{\cal W}_*^{\text{RR}}
\ &= \ 
- \frac{24 \Delta^{\text{RR}}}{(\lambda^{\text{RR}})^3}
\Big(
36 (m_{\text{RR}})^3 + 36 (m_{\text{RR}}^0)^2 e_{\text{RR}0}
- 3 m_{\text{RR}} \lambda^{\text{RR}} 
- 4 i \, \text{sign}(\lambda^{\text{RR}}) m_{\text{RR}}^0 \sqrt{ 3 \Delta^{\text{RR}} }
\Big)
\, . \label{IIA-AP-WR_sol}
\end{align}
\esubeq
These expressions are quite sensitive to signs of the discriminants 
$\Delta^{\text{RR}}$ and $\lambda^{\text{RR}}$.


If $\Delta^{\text{RR}}$
is positive, $\lambda^{\text{RR}}$ is always positive. 
Under this condition we find that the expression $t_*$ (\ref{IIA-AP-WR-t_sol}) 
becomes a consistent solution and that the superpotential does
not vanish:
\bsubeq \label{IIA-WR-single-D>0-sol}
\begin{align}
t_* \ &= \ 
\frac{6(3 m_{\text{RR}}^0 e_{\text{RR}0} + m_{\text{RR}} e_{\text{RR}})}{\lambda^{\text{RR}}}
- \frac{2 i \sqrt{3 \Delta^{\text{RR}}}}{\lambda^{\text{RR}}}
\, , \\
{\cal W}_*^{\text{RR}}
\ &= \ 
- \frac{24 \Delta^{\text{RR}}}{(\lambda^{\text{RR}})^3}
\Big(
36 (m_{\text{RR}})^3 + 36 (m_{\text{RR}}^0)^2 e_{\text{RR}0}
- 3 m_{\text{RR}} \lambda^{\text{RR}} 
- 4 i \, m_{\text{RR}}^0 \sqrt{ 3 \Delta^{\text{RR}} }
\Big)
\, .
\end{align}
\esubeq
Here we chose the minus sign in front of $t_{2*}$ in order that the
K\"{a}hler potential 
$K_+ = - \log [ - i (t_* - \ol{t}_*)^3]$ is well-defined.


If $\Delta^{\text{RR}}$ vanishes,
$\lambda^{\text{RR}}$ is non-negative. 
However, if $\lambda^{\text{RR}}$ also vanishes,
$t_*$ and ${\cal W}_*^{\text{RR}}$ become singular. 
This is forbidden.
In the case of positive $\lambda^{\text{RR}}$,
$t_*$ is real and ${\cal W}_*^{\text{RR}}$ vanishes.
Although this point is harmless as far as the equation 
$D_t {\cal W}^{\text{RR}} = 0$ is concerned,
it should not be chosen as an admissible supersymmetric solution, 
because the metric and the curvature tensor become singular: 
\begin{align}
K_{t \ol{t}} \ &= \ 
- \frac{3}{(t - \ol{t})^2}
\, , \ls
R^t{}_{tt\ol{t}} \ = \ 
\frac{2}{(t - \ol{t})^2}
\, . \label{curve-WR-real-singular}
\end{align}
We conclude that if the discriminant $\Delta^{\text{RR}}$ vanishes,
there are no physical solutions.


If $\Delta^{\text{RR}}$ is negative,
$t_{2*}$ in (\ref{IIA-AP-WR-t_sol}) is ill-defined.
This implies that there are no consistent solutions of 
the equation $D_t {\cal W}^{\text{RR}}|_* = 0$, even though 
$\lambda^{\text{RR}}$ is not restricted.


\subsubsection{Solutions of ${\cal W}^{\rm RR} = 0$}

Here we look for a consistent solution which satisfies the equation 
${\cal W}_*^{\text{RR}} = 0$.
In this consideration it is also useful to classify 
physical solutions in terms of the discriminant 
$\Delta^{\text{RR}}$ (\ref{IIA-WRR-discrim}).


If $\Delta^{\text{RR}}$ is positive, 
there are three distinct real roots $(e_1, e_2, e_3)$
of the equation ${\cal W}^{\text{RR}}  = 0$.
The superpotential and its K\"{a}hler covariant derivative are rewritten as
\bsubeq \label{IIA-AP-WR0-D>0}
\begin{align}
{\cal W}^{\text{RR}}
\ &= \ 
m_{\text{RR}}^0 (t - e_1) (t - e_2) (t - e_3)
\, , \ls e_1, e_2, e_3 \ \in \ {\mathbb R}
\, , \\
D_t {\cal W}^{\text{RR}}
\ &= \ 
- \frac{{\cal W}^{\text{RR}}}{t - \ol{t}}
\left( \frac{\ol{t} - e_1}{t - e_1}
+ \frac{\ol{t} - e_2}{t - e_2}
+ \frac{\ol{t} - e_3}{t - e_3}
\right)
\, .
\end{align}
The three real roots $e_i$ are related to the Ramond-Ramond flux charges:
\begin{align}
3 m_{\text{RR}} \ &= \ m_{\text{RR}}^0 \big( e_1 + e_2 + e_3 \big)
\, , \ \ \ 
e_{\text{RR}} \ = \ m_{\text{RR}}^0 \big( e_1 e_2 + e_2 e_3 + e_3 e_1 \big)
\, , \ \ \
e_{\text{RR}0} \ = \ - m_{\text{RR}}^0 e_1 e_2 e_3 
\, . \label{IIA-AP-WR0-sol_charges1}
\end{align}
We find a non-zero value of the covariant derivative at the
points $t_* = e_i$. For instance, the value at $t_* = e_1$ is
\begin{align}
D_t {\cal W}^{\text{RR}} \big|_{t_* = e_1}
\ &= \ 
- 3 m_{\text{RR}}^0 (e_1 - e_2) (e_1 - e_3)
\ \neq \ 0
\, .
\end{align}
\esubeq
This value itself is finite. However, the K\"{a}hler metric
and the curvature (\ref{curve-WR-real-singular}) become singular. 
Then we cannot choose this solution as an attractor point.
The other two zeros $e_2$ and $e_3$ give the same
situations. Thus there are no finite solutions of 
${\cal W}^{\text{RR}} = 0$ if $\Delta^{\text{RR}}$ is positive. 


If $\Delta^{\text{RR}}$ vanishes,
$\lambda^{\text{RR}}$ is non-negative.
When $\lambda^{\text{RR}}$ is positive,
the equation ${\cal W}^{\text{RR}} =0$ has 
two coincident real roots $e_1$ and a distinct real root $e_2$.
When $\lambda^{\text{RR}}$ vanishes, 
the three roots coincide with each other.
In both cases the superpotential and its
covariant derivative are 
\bsubeq \label{IIA-AP-WR0-D=0}
\begin{align}
{\cal W}^{\text{RR}}
\ &= \ 
m_{\text{RR}}^0 (t - e_1)^2 (t - e_2)
\, , \ls e_1, e_2 \ \in \ {\mathbb R}
\, , \\
D_t {\cal W}^{\text{RR}}
\ &= \ 
- \frac{{\cal W}^{\text{RR}}}{t - \ol{t}}
\left( \frac{2 (\ol{t} - e_1)}{t - e_1}
+ \frac{\ol{t} - e_2}{t - e_2}
\right)
\, .
\end{align}
The relations among the flux charges and the roots are
\begin{align}
3 m_{\text{RR}} \ &= \ m_{\text{RR}}^0 \big( 2 e_1 + e_2 \big)
\, , \ls
e_{\text{RR}} \ = \ m_{\text{RR}}^0 \big( (e_1)^2 + 2 e_1 e_2 \big)
\, , \ls
e_{\text{RR}0} \ = \ - m_{\text{RR}}^0 (e_1)^2 e_2 
\, . \label{IIA-AP-WR0-sol_charges2}
\end{align}
We find that the covariant derivatives of the superpotential
vanish at the points $t_* = e_i$:
\begin{align}
D_t {\cal W}^{\text{RR}} \big|_{t_* = e_1}
\ &= \ 
0
\, , \ls
D_t {\cal W}^{\text{RR}} \big|_{t_* = e_2}
\ = \ 
0
\, .
\end{align}
\esubeq
These values are finite. However, the K\"{a}hler metric
and the curvature (\ref{curve-WR-real-singular}) 
become singular in the same reason as in $\Delta^{\text{RR}} > 0$.
They are inadmissible to physical solutions.


If $\Delta^{\text{RR}}$ is negative, 
the equation ${\cal W}^{\text{RR}} =0$ has 
one real root $e_1$ and a pair of complex roots $(\alpha, \ol{\alpha})$.
Then the superpotential and its covariant derivative are rewritten as 
\bsubeq \label{IIA-AP-WR0-D<0}
\begin{align} 
{\cal W}^{\text{RR}}
\ &= \ 
m_{\text{RR}}^0 (t - e_1) (t - \alpha) (t - \ol{\alpha})
\, , \ls e_1 \ \in \ {\mathbb R}
\, , \ \ \ \alpha \ \in \ {\mathbb C}
\, , \\
D_t {\cal W}^{\text{RR}}
\ &= \ 
- \frac{{\cal W}^{\text{RR}}}{t - \ol{t}}
\left( \frac{\ol{t} - e_1}{t - e_1}
+ \frac{\ol{t} - \alpha}{t - \alpha}
+ \frac{\ol{t} - \ol{\alpha}}{t - \ol{\alpha}}
\right)
\, .
\end{align}
\esubeq
The three roots are related to the flux charges:
\begin{gather}
3 m_{\text{RR}} \ = \ m_{\text{RR}}^0 \big( e_1 + \alpha + \ol{\alpha} \big)
\, , \ \ \ 
e_{\text{RR}} \ = \ m_{\text{RR}}^0 
\big( e_1 ( \alpha + \ol{\alpha} ) + |\alpha|^2 \big)
\, , \ls 
e_{\text{RR}0} \ = \ - m_{\text{RR}}^0 e_1 |\alpha|^2 
\, . \label{IIA-AP-WR0-sol_charges3}
\end{gather}
The solutions are explicitly given by
\bsubeq \label{IIA-AP-WR0-sol_charges33}
\begin{align}
e_1 \ &= \ 
- \frac{1}{m_{\text{RR}}^0}
\Big( -3 m_{\text{RR}} + 2 m_{\text{RR}}^0 ({\rm Re} \, \alpha) 
\Big)
\, , \\
({\rm Re} \, \alpha)
\ &= \
\frac{\lambda^{\text{RR}} + (F_{\text{RR}})^{2/3} 
+ 12 m_{\text{RR}} (F_{\text{RR}})^{1/3}}{12 m_{\text{RR}}^0 (F_{\text{RR}})^{1/3}}
\ls \text{(if $F_{\text{RR}} > 0$)}
\, , \\
\text{or} \ \ \ 
({\rm Re} \, \alpha)
\ &= \ 
- \frac{1}{24 m_{\text{RR}}^0 (F_{\text{RR}})^{1/3}}
\Big( \big( \lambda^{\text{RR}} + (F_{\text{RR}})^{2/3} \big) 
\pm \sqrt{3} i \big( \lambda^{\text{RR}} - (F_{\text{RR}})^{2/3} \big) 
- 24 m_{\text{RR}} (F_{\text{RR}})^{1/3} \Big)
\nn \\
\ &= \
\frac{\lambda^{\text{RR}} + (G_{\text{RR}})^{2/3} 
+ 12 m_{\text{RR}} (G_{\text{RR}})^{1/3}}{12 m_{\text{RR}}^0 (G_{\text{RR}})^{1/3}}
\ls \text{(if $F_{\text{RR}} = - G_{\text{RR}} < 0$)}
\, , \\
({\rm Im} \, \alpha)^2
\ &= \ 
\frac{1}{m_{\text{RR}}^0}
\Big(
e_{\text{RR}}
- 6 m_{\text{RR}} ({\rm Re} \, \alpha)
+ 3 m_{\text{RR}}^0 ({\rm Re} \, \alpha)^2
\Big)
\, , \\
F_{\text{RR}} \ &= \ 
108 (m_{\text{RR}}^0)^2 e_{\text{RR}0} 
+ 12 m_{\text{RR}}^0 \sqrt{- 3 \Delta^{\text{RR}}} 
+ 108 (m_{\text{RR}})^3 
- 9 \lambda^{\text{RR}} m_{\text{RR}}
\, .
\end{align}
\esubeq
Note that $F_{\text{RR}}$ cannot vanish otherwise $t_* = \alpha$ goes to infinity.
In order that the above expressions 
provide a solution of ${\cal W}_*^{\text{RR}} =0$ and 
$D_t {\cal W}^{\text{RR}} |_*\neq 0$,
the square of the imaginary part of $\alpha$ has to be positive definite:
\begin{align}
3 m_{\text{RR}}^0 ({\rm Re} \, \alpha)^2 
- 6 m_{\text{RR}} ({\rm Re} \, \alpha)
+ e_{\text{RR}}
\ > \ 0
\, . \label{IIA-AP-WR0_neg_cond}
\end{align}
The discriminant of the function of $({\rm Re} \, \alpha)$ 
in the left-hand side is nothing but $\lambda^{\text{RR}}$.
If this is non-negative, 
there exist the following points where $({\rm Im} \, \alpha)$ vanishes:
\begin{align}
({\rm Re} \, \alpha)
\ &= \ 
\frac{1}{6 m_{\text{RR}}^0} \big( 6 m_{\text{RR}} \pm \sqrt{\lambda^{\text{RR}}} \big)
\, .
\end{align}
However, this is inconsistent with $\Delta^{\text{RR}} < 0$ 
that gives one real and a pair of complex zeros.
Then we find that $\lambda^{\text{RR}} < 0$ is necessary 
to obtain a solution of ${\cal W}_*^{\text{RR}} = 0$ with
$D_t {\cal W}^{\text{RR}} |_* \neq 0$.
Since the root $t_* = e_1$ gives singular curvature,
the consistent solution is only given by $t_* = \alpha$.


\subsection{(Non)geometric flux superpotential}
\label{IIA-single-WQ-generic}

In this subsection we investigate features of the 
(non)geometric flux superpotential.
Since the function ${\cal W}^{\cal Q}$ is similar to ${\cal W}^{\text{RR}}$,
we can evaluate this sector in a parallel way as in the previous subsection.
First we look for a solution of
$D_t {\cal W}^{\cal Q} = 0$.
Next we analyze a condition 
${\cal W}^{\cal Q} = 0$ by using the discriminants 
$\Delta^{\cal Q}$ and $\lambda^{\cal Q}$ in (\ref{IIA-WQ-discrim}).

\subsubsection{Solutions of $D_t {\cal W}^{\cal Q} = 0$}

Let us investigate  consistent conditions to satisfy 
the equation $D_t {\cal W}^{\cal Q} = 0$.
We formally describe a solution of 
$D_t {\cal W}^{\cal Q} = 0$ as follows:
\bsubeq \label{IIA-AP-DWQ0}
\begin{align}
t_* \ &\equiv \ t_{1*} + i t_{2*}
\ = \  
- \frac{6(3 p_0{}^0 e_{00} + p_0 e_{0})}{\lambda^{\cal Q}}
\pm \frac{2 i \sqrt{3 \Delta^{\cal Q}}}{\lambda^{\cal Q}}
\, . \label{IIA-AP-WQ-t_sol}
\end{align}
The superpotential at this point is given by
\begin{align}
{\cal W}_*^{\cal Q}
\ &= \ 
- \frac{24 \Delta^{\cal Q}}{(\lambda^{\cal Q})^3}
\Big(
36 (p_0)^3 - 36 (p_0{}^0)^2 e_{00}
- 3 p_0 \lambda^{\cal Q} 
- 4 i \, \text{sign}(\lambda^{\cal Q}) p_0{}^0 \sqrt{ 3 \Delta^{\cal Q} }
\Big)
\, . \label{IIA-AP-WQ_sol}
\end{align}
\esubeq
Consistency of the above formal expression is evaluated in terms
of the discriminants $\Delta^{\cal Q}$ and $\lambda^{\cal Q}$ 
as in the previous subsection.


If $\Delta^{\cal Q}$
is positive, $\lambda^{\cal Q}$ is always positive. 
Under this condition we find that $t_*$ (\ref{IIA-AP-WQ-t_sol}) 
becomes a consistent solution with non-vanishing superpotential:
\bsubeq \label{IIA-WQ-single-D>0-sol}
\begin{align}
t_* \ &= \ 
- \frac{6(3 p_0{}^0 e_{00} + p_0 e_{0})}{\lambda^{\cal Q}}
- \frac{2 i \sqrt{3 \Delta^{\cal Q}}}{\lambda^{\cal Q}}
\, , \\
{\cal W}_*^{\cal Q}
\ &= \ 
- \frac{24 \Delta^{\cal Q}}{(\lambda^{\cal Q})^3}
\Big(
36 (p_0)^3 - 36 (p_0{}^0)^2 e_{00}
- 3 p_0 \lambda^{\cal Q} 
- 4 i \, p_0{}^0 \sqrt{ 3 \Delta^{\cal Q} }
\Big)
\, .
\end{align}
\esubeq
Here we have already chose the negative sign in front of $t_{2*}$
to realize a well-defined K\"{a}hler potential.
We find the K\"{a}hler metric is non-degenerated and the curvature
is finite.


If $\Delta^{\cal Q}$ vanishes,
$\lambda^{\cal Q}$ is non-negative. 
However if $\lambda^{\cal Q}$ is zero, 
$t_{1*}$ in (\ref{IIA-AP-WQ-t_sol}) and ${\cal W}_*^{\cal Q}$ (\ref{IIA-AP-WQ_sol})
are ill-defined. Then only the positive $\lambda^{\cal Q}$ is allowed.
In this case,
$t_*$ is reduced to a real value and ${\cal W}_*^{\cal Q}$ vanishes.
It cannot be chosen as a physical solution to realize a well-defined supersymmetric
solution, because the curvature tensor (\ref{curve-WR-real-singular})
goes to infinity.
We conclude that there are no admissible solutions of $D_t {\cal W}^{\cal Q} = 0$
if $\Delta^{\cal Q}$ vanishes.


If $\Delta^{\cal Q}$ is negative,
the expression $t_{2*}$ in (\ref{IIA-AP-WQ-t_sol}) becomes ill-defined.
This implies that there are no consistent solutions of 
the equation $D_t {\cal W}^{\cal Q}|_* = 0$, even though 
the discriminant $\lambda^{\cal Q}$
is not restricted.


\subsubsection{Solutions of ${\cal W}^{\cal Q} = 0$}

Here we look for a consistent solution of the equation ${\cal W}_*^{\cal Q} = 0$.
If $\Delta^{\cal Q}$ is positive or zero, 
there are no consistent solutions to realize supersymmetric vacua
as in the previous subsection.
Then we focus on the case of the negative valued $\Delta^{\cal Q}$.
In this case, the equation ${\cal W}^{\cal Q} =0$ has 
one real root $e_1$ and a pair of complex roots $(\alpha, \ol{\alpha})$.
The superpotential and its covariant derivative are written as
\bsubeq \label{IIA-AP-WQ0-D<0}
\begin{align} 
{\cal W}^{\cal Q}
\ &= \ 
p_0{}^0 (t - e_1) (t - \alpha) (t - \ol{\alpha})
\, , \ls e_1 \ \in \ {\mathbb R}
\, , \ \ \ \alpha \ \in \ {\mathbb C}
\, , \\
D_t {\cal W}^{\cal Q}
\ &= \ 
- \frac{{\cal W}^{\cal Q}}{t - \ol{t}}
\left( \frac{\ol{t} - e_1}{t - e_1}
+ \frac{\ol{t} - \alpha}{t - \alpha}
+ \frac{\ol{t} - \ol{\alpha}}{t - \ol{\alpha}}
\right)
\, .
\end{align}
\esubeq
The three roots are related to the flux charges:
\begin{gather}
3 p_0 \ = \ p_0{}^0 \big( e_1 + \alpha + \ol{\alpha} \big)
\, , \ \ \ 
e_0 \ = \ - p_0{}^0 \big( e_1 ( \alpha + \ol{\alpha} ) + |\alpha|^2 \big)
\, , \ls 
e_{00} \ = \ p_0{}^0 e_1 |\alpha|^2 
\, . \label{IIA-AP-WQ0-sol_charges3}
\end{gather}
The solutions are given by 
\bsubeq \label{IIA-AP-WQ0-sol_charges33}
\begin{align}
e_1 \ &= \ 
- \frac{1}{p_0{}^0}
\Big( -3 p_0 + 2 p_0{}^0 ({\rm Re} \, \alpha) 
\Big)
\, , \\
({\rm Re} \, \alpha)
\ &= \
\frac{\lambda^{\cal Q} + (F_{\cal Q})^{2/3} 
+ 12 p_0 (F_{\cal Q})^{1/3}}{12 p_0{}^0 (F_{\cal Q})^{1/3}}
\ls \text{(if $F_{\cal Q} > 0$)}
\, , \\
\text{or} \ \ \ 
({\rm Re} \, \alpha)
\ &= \ 
- \frac{1}{24 p_0{}^0 (F_{\cal Q})^{1/3}}
\Big( \big( \lambda^{\cal Q} + (F_{\cal Q})^{2/3} \big) 
\pm \sqrt{3} i \big( \lambda^{\cal Q} - (F_{\cal Q})^{2/3} \big) 
- 24 p_0 (F_{\cal Q})^{1/3} \Big)
\nn \\
\ &= \
\frac{\lambda^{\cal Q} + (G_{\cal Q})^{2/3} 
+ 12 p_0 (G_{\cal Q})^{1/3}}{12 p_0{}^0 (G_{\cal Q})^{1/3}}
\ls \text{(if $F_{\cal Q} = - G_{\cal Q} < 0$)}
\, , \\
({\rm Im} \, \alpha)^2
\ &= \ 
\frac{1}{p_0{}^0}
\Big(
- e_0
- 6 p_0 ({\rm Re} \, \alpha)
+ 3 p_0{}^0 ({\rm Re} \, \alpha)^2
\Big)
\, , \\
F_{\cal Q} \ &= \ 
- 108 (p_0{}^0)^2 e_{00} 
+ 12 p_0{}^0 \sqrt{- 3 \Delta^{\cal Q}} 
+ 108 (p_0)^3 
- 9 \lambda^{\cal Q} p_0
\, .
\end{align}
\esubeq
Note that $F_{\cal Q}$ is non-zero otherwise $t_* = \alpha$ goes to infinity.
Since we have already assumed $p_0{}^0 > 0$,
the following inequality should be imposed:
\begin{align}
3 p_0{}^0 ({\rm Re} \, \alpha)^2 
- 6 p_0 ({\rm Re} \, \alpha)
- e_0
\ > \ 0
\, . \label{IIA-AP-WQ0_neg_cond}
\end{align}
The discriminant of the function of $({\rm Re} \, \alpha)$ 
in the left-hand side is nothing but $\lambda^{\cal Q}$.
If this is non-negative, there exist the following points where 
$({\rm Im} \, \alpha)$ vanishes:
\begin{align}
({\rm Re} \, \alpha)
\ &= \ \frac{1}{6 p_0{}^0} \big( 6 p_0 \pm \sqrt{\lambda^{\cal Q}} \big)
\, .
\end{align}
However, this is inconsistent with the condition 
$\Delta^{\cal Q} < 0$ which gives one real and a pair of complex
zeros.
Then $\lambda^{\cal Q} < 0$ is necessary 
to obtain a solution of ${\cal W}_*^{\cal Q} = 0$ with
$D_t {\cal W}^{\cal Q} |_* \neq 0$.
Since $t_* = e_1$ gives ill-defined curvature, 
the consistent solution is only given by $t_* = \alpha$.


\subsection{Supersymmetric vacua}

We have already studied various situations when
the superpotentials ${\cal W}^{\text{RR}}$ and ${\cal W}^{\cal Q}$ and/or
their covariant derivatives $D_t {\cal W}^{\text{RR}}$ and 
$D_t {\cal W}^{\cal Q}$ have zeros.
The signs of the discriminants of the superpotentials 
characterize admissible solutions.
Here we classify supersymmetric flux attractor vacua 
which satisfy (\ref{IIA-AP-BBW2}).

Consider the case that 
both the two discriminants $\Delta^{\text{RR}}$ and $\Delta^{\cal Q}$ are positive.
There exists a solution which satisfies
$D_t {\cal W}^{\text{RR}} = 0$, $D_t {\cal W}^{\cal Q} = 0$,
${\cal W}^{\text{RR}} \neq 0$ and ${\cal W}^{\cal Q} \neq 0$. 
Here obtain the following equations from (\ref{IIA-WR-single-D>0-sol})
and (\ref{IIA-WQ-single-D>0-sol}):
\bsubeq
\begin{align}
D_t {\cal W} \big|_*
\ &= \ 
D_t {\cal W}^{\text{RR}} \big|_* 
+ U_* D_t {\cal W}^{\cal Q} \big|_* 
\ = \ 
0 
\, , \ls
D_t {\cal W}^{\text{RR}} \big|_* 
\ = \ 
D_t {\cal W}^{\cal Q} \big|_* 
\ = \ 
0 
\, , \\
D_U {\cal W} \big|_*
\ &= \ 
\frac{1}{{\rm Im} U} 
\Big( {\cal W}_*^{\text{RR}} + {\rm Re} U_* {\cal W}_*^{\cal Q} \Big)
\ = \ 
0 
\, , \\
{\cal W}_*
\ &= \ 
{\cal W}_*^{\text{RR}} + U_* {\cal W}_*^{\cal Q}
\ = \ 
i {\rm Im} U_*
{\cal W}_*^{\cal Q}
\, , \\
t_*^{\text{RR}} \ &= \ 
\frac{6(3 m_{\text{RR}}^0 e_{\text{RR}0} + m_{\text{RR}} e_{\text{RR}})}{\lambda^{\text{RR}}}
- \frac{2 i \sqrt{3 \Delta^{\text{RR}}}}{\lambda^{\text{RR}}}
\, , \\
t_*^{\cal Q} \ &= \ 
- \frac{6(3 p_0{}^0 e_{00} + p_0 e_{0})}{\lambda^{\cal Q}}
- \frac{2 i \sqrt{3 \Delta^{\cal Q}}}{\lambda^{\cal Q}}
\, , \\
{\cal W}_*^{\text{RR}}
\ &= \ 
- \frac{24 \Delta^{\text{RR}}}{(\lambda^{\text{RR}})^3}
\Big(
36 (m_{\text{RR}})^3 + 36 (m_{\text{RR}}^0)^2 e_{\text{RR}0}
- 3 m_{\text{RR}} \lambda^{\text{RR}} 
- 4 i \, m_{\text{RR}}^0 \sqrt{ 3 \Delta^{\text{RR}} }
\Big)
\, , \\
{\cal W}_*^{\cal Q}
\ &= \ 
- \frac{24 \Delta^{\cal Q}}{(\lambda^{\cal Q})^3}
\Big(
36 (p_0)^3 - 36 (p_0{}^0)^2 e_{00}
- 3 p_0 \lambda^{\cal Q} 
- 4 i \, p_0{}^0 \sqrt{ 3 \Delta^{\cal Q} }
\Big)
\, . 
\end{align}
\esubeq
Since the two solutions $t_*^{\text{RR}}$ and $t_*^{\cal Q}$ have to
coincide with each other, we find a non-trivial relation:
\begin{align}
\frac{3 m_{\text{RR}}^0 e_{\text{RR}0} + m_{\text{RR}} e_{\text{RR}}}{\lambda^{\text{RR}}}
\ &= \ 
- \frac{3 p_0{}^0 e_{00} + p_0 e_{0}}{\lambda^{\cal Q}}
\, , \ls
\frac{\sqrt{\Delta^{\text{RR}}}}{\lambda^{\text{RR}}}
\ = \ 
\frac{\sqrt{\Delta^{\cal Q}}}{\lambda^{\cal Q}}
\, .
\end{align}
We can fix only the real part of the variable $U$ by
\begin{align}
{\rm Re} U_* \ &= \ 
- \frac{{\cal W}_*^{\text{RR}}}{{\cal W}_*^{\cal Q}}
\, ,
\end{align}
whilst the imaginary part remains unfixed.
This indicates that the dilaton (\ref{IIAdilatonN=1}) is not fixed.
The value of the superpotential
${\cal W}_*$ also contains ${\rm Im} U$.
However, this does not explicitly appear in the cosmological constant
$\Lambda = - 3 \, \e^K |{\cal W}_*|^2$:
\begin{align}
- 3 \, \e^{K} |{\cal W}_*|^2
\ &= \ 
\frac{3}{2 (t_2^{\cal Q})^3} 
\frac{1}{[{\rm Re} ({\cal C} {\cal G}_0)]^2}
| {\cal W}_*^{\cal Q} |^2
\ = \
- \frac{4}{[{\rm Re} ({\cal C} {\cal G}_0)]^2} 
\sqrt{\frac{\Delta^{\cal Q}}{3}}
\, . \label{ex1_cos_const}
\end{align}
The value ${\rm Re} ({\cal C} {\cal G}_0)$, which should be
non-zero to realize a well-defined K\"{a}hler potential
(\ref{IIAKahlerN=1}), is not fixed 
by the attractor equations, either.
However this should be very large
under the supergravity approximation: 
The exponent of the expectation value of the dilaton
gives the sting coupling constant.
This should be very small.
This restriction imposes that the compensator ${\cal C}$ (\ref{IIAPiC}) is
very large. 
Then the cosmological constant (\ref{ex1_cos_const})
becomes very small. 
This solution realizes a supersymmetric AdS vacuum.
The stability of the system has already been guaranteed by
\cite{deCarlos:2005kh} in a generic form, where 
all mass eigenvalues satisfy the Breitenlohner-Freedman criterion
\cite{Breitenlohner:1982bm}.
This result differs from that of \cite{Anguelova:2008fm} where
only the Minkowski vacua is realized. 
This difference comes from the introduction of the Ramond-Ramond flux
charges. We will come back to this issue in later sections. 

Next, let us consider the case that 
both of $\Delta^{\text{RR}}$ and $\Delta^{\cal Q}$ are negative.
There exists another attractor point which satisfies 
${\cal W}^{\text{RR}} = 0$, ${\cal W}^{\cal Q} = 0$,
$D_t {\cal W}^{\text{RR}} \neq 0$ and $D_t {\cal W}^{\cal Q} \neq 0$.
We can see a non-trivial relation between the Ramond-Ramond flux
charges and the (non)geometric flux charges via the equations 
${\cal W}^{\text{RR}} = 0$ and ${\cal W}^{\cal Q} = 0$.
The former gives a solution
$t_* = \alpha^{\text{RR}} (e_{\text{RR}0}, e_{\text{RR}},
m_{\text{RR}}, m_{\text{RR}}^0)$ 
in (\ref{IIA-AP-WR0-sol_charges33}), 
while the latter yields 
$t_* = \alpha^{\cal Q} (e_{00}, e_0, p_0, p_0{}^o)$ 
in (\ref{IIA-AP-WQ0-sol_charges33}).
These two solutions have to coincide with each other:
\bsubeq
\begin{align}
\alpha^{\text{RR}} \ &= \ 
\alpha^{\cal Q} 
\, , \\
{\rm Re} \, \alpha^{\text{RR}} 
\ &= \ 
\frac{\lambda^{\text{RR}} + (F_{\text{RR}})^{2/3} 
+ 12 m_{\text{RR}} (F_{\text{RR}})^{1/3}}{12 m_{\text{RR}}^0 (F_{\text{RR}})^{1/3}}
\, , \\
{\rm Re} \, \alpha^{\cal Q} 
\ &= \ 
\frac{\lambda^{\cal Q} + (F_{\cal Q})^{2/3} 
+ 12 p_0 (F_{\cal Q})^{2/3}}{12 p_0{}^0 (F_{\cal Q})^{1/3}}
\, , \\
({\rm Im} \, \alpha^{\text{RR}})^2
\ &= \ 
\frac{1}{m_{\text{RR}}^0} \Big(
e_{\text{RR}} - 6 m_{\text{RR}} ({\rm Re} \, \alpha^{\text{RR}})
+ 3 m_{\text{RR}}^0 ({\rm Re} \, \alpha^{\text{RR}})
\Big)
\, , \\
({\rm Im} \, \alpha^{\cal Q})^2
\ &= \ 
\frac{1}{p_0{}^0} \Big(
- e_0 - 6 p_0 ({\rm Re} \, \alpha^{\cal Q})
+ 3 p_0{}^0 ({\rm Re} \, \alpha^{\cal Q})^2
\Big)
\, , \\
e_1^{\text{RR}}
\ &= \ 
- \frac{1}{m_{\text{RR}}^0} \Big( - 3 m_{\text{RR}} + 2 m_{\text{RR}}^0
({\rm Re} \, \alpha^{\text{RR}}) \Big)
\, , \\
e_1^{\cal Q}
\ &= \ 
- \frac{1}{p_0{}^0} \Big( - 3 p_0 + 2 p_0{}^0 ({\rm Re} \, \alpha^{\cal Q}) \Big)
\, , \\
F_{\text{RR}}
\ &= \ 
108 (m_{\text{RR}}^0)^2 e_{\text{RR}0}
+ 12 m_{\text{RR}}^0 \sqrt{- 3 \Delta^{\text{RR}}} 
+ 108 (m_{\text{RR}})^3 - 9 \lambda^{\text{RR}} m_{\text{RR}}
\, , \\
F_{\cal Q}
\ &= \ 
- 108 (p_0{}^0)^2 e_{00} 
+ 12 p_0{}^0 \sqrt{- 3 \Delta^{\cal Q}} 
+ 108 (p_0)^3 - 9 \lambda^{\cal Q} p_0
\, .
\end{align}
\esubeq
We can stabilize the variable $U$ in the following way:
\bsubeq \label{IIA-case3sol-D0D2D6}
\begin{align}
U_* 
\ &= \ 
- \frac{D_t {\cal W}^{\text{RR}} |_{t_* = \alpha}}
{D_t {\cal W}^{\cal Q} |_{t_* = \alpha}}
\, , \\
D_t {\cal W}^{\text{RR}} \big|_{t_* = \alpha}
\ &= \ 
- 2 i \, m_{\text{RR}}^0 ({\rm Im} \, \alpha^{\text{RR}})
\Big[
3 \Big( \frac{m_{\text{RR}}}{m_{\text{RR}}^0} - ({\rm Re} \, \alpha^{\text{RR}}) \Big)
- i \, ({\rm Im} \, \alpha^{\text{RR}}) 
\Big]
\, , \\
D_t {\cal W}^{\cal Q} \big|_{t_* = \alpha}
\ &= \ 
- 2i \, p_0{}^0 ({\rm Im} \, \alpha^{\cal Q}) 
\Big[
3 \Big( \frac{p_0}{p_0{}^0} - ({\rm Re} \, \alpha^{\cal Q}) \Big)
- i \, ({\rm Im} \, \alpha^{\cal Q}) 
\Big]
\, , 
\end{align}
\esubeq
where we used ${\rm Im} U \neq 0$ because of finiteness of the
curvature tensor $R^U{}_{UU\ol{U}}$.
The vanishing superpotential sets the cosmological constant to be zero.
Then a supersymmetric Minkowski vacuum is realized.
This configuration is interpreted that
the internal space $\mathscr{M}$ is reduced to a parallelizable twisted torus
\cite{Grana:2006kf}.

We discuss other situations:
(i) There are no attractor solutions to satisfy the equations (\ref{IIA-AP-BBW2}) if 
the relative signs of the two discriminants are different;
i.e., $\Delta^{\text{RR}} \cdot \Delta^{\cal Q} < 0$.
(ii) Apart from the attractor solutions where the moduli are stabilized,
there exist non-attractor solutions which satisfy
the supersymmetry condition (\ref{IIA-AP-BBW2}). 
Due to the lack of the number of equations, however,
the moduli $t$ and $U$ are not fixed at all.
These solutions do not provide vanishing superpotentials.
Then the vacua are realized as AdS spaces.


\section{Example 2: a model without Ramond-Ramond flux charges}
\label{section-RR0}

In this section we study a model without the Ramond-Ramond flux charges
$e_{\text{RR} \check{A}} = 0 = m_{\text{RR}}^{\check{A}}$.
The total superpotential ${\cal W}$ and its covariant
derivatives are reduced to 
\bsubeq
\begin{align}
{\cal W} \ &= \ U {\cal W}^{\cal Q}
\, , \\
{\cal W}^{\cal Q}
\ &= \ 
- e_{00} - e_{0 \check{a}} t^{\check{a}}
- 3 p_0{}^{\check{c}} D_{\check{a} \check{b} \check{c}} \, t^{\check{a}} t^{\check{b}} 
+ p_0{}^0 D_{\check{a} \check{b} \check{c}} \, 
t^{\check{a}} t^{\check{b}} t^{\check{c}}
\, , \\
D_{\check{a}} {\cal W}
\ &= \ 
U D_{\check{a}} {\cal W}^{\cal Q}
\, , \ls
D_U {\cal W}
\ = \ 
i \, \frac{{\rm Re} U}{{\rm Im} U}
{\cal W}^{\cal Q}
\, .
\end{align}
\esubeq 
We imposed ${\rm Im} \, U \neq 0$.
In supersymmetric solutions, the following equations have to be
satisfied:
\bsubeq \label{IIA-RR0-SUSY_sol_cond}
\begin{alignat}{3}
D_{\check{a}} {\cal W} \ &= \ 
0 & \ls &\leftrightarrow & \ls
D_{\check{a}} {\cal W}^{\cal Q}
\ &= \ 0
\, , \\
D_U {\cal W} \ &= \ 
0 & \ls &\leftrightarrow & \ls
{\rm Re} U
{\cal W}^{\cal Q}
\ &= \ 
0 \, . 
\end{alignat}
\esubeq
In the single modulus model as in section \ref{IIA-single-WQ-generic},
we obtain a solution (\ref{IIA-WQ-single-D>0-sol})
consistent with (\ref{IIA-RR0-SUSY_sol_cond}):
\bsubeq
\begin{align}
\Delta^{\cal Q}
\ &= \ 
-27 (p_0{}^0 e_{00})^2
- 54 p_0{}^0 e_{00} p_0 e_0 
+ 9 (p_0 e_0)^2 
- 108 (p_0)^3 e_{00}
+ 4 p_0{}^0 (e_0)^3
\ > \ 0
\, , \\
\lambda^{\cal Q}
\ &= \ 
12 \big( 3 (p_0)^2 + p_0{}^0 e_0 \big) 
\ > \ 0
\, , \\
t_* \ &= \ 
- \frac{6(3 p_0{}^0 e_{00} + p_0 e_{0})}{\lambda^{\cal Q}}
- \frac{2 i \sqrt{3 \Delta^{\cal Q}}}{\lambda^{\cal Q}}
\, , \\
{\cal W}_*^{\cal Q}
\ &= \ 
- \frac{24 \Delta^{\cal Q}}{(\lambda^{\cal Q})^3}
\Big(
36 (p_0)^3 - 36 (p_0{}^0)^2 e_{00}
- 3 p_0 \lambda^{\cal Q} 
- 4 i \, p_0{}^0 \sqrt{ 3 \Delta^{\cal Q} }
\Big)
\, , \\
{\rm Re} U_* \ &= \ 0
\, .
\end{align}
\esubeq
Here we chose that $t_{2*}$ is negative in order
that the K\"{a}hler potential is well-defined. 
The scalar potential at this point is described as
\begin{align}
V_* \ &= \ 
- \frac{4}{[{\rm Re} ({\cal C} {\cal G}_0)]^2} 
\sqrt{\frac{\Delta^{\cal Q}}{3}}
\, . \label{cos-const-2}
\end{align}
In this model the attractor equations (\ref{IIA-RR0-SUSY_sol_cond})
can fix only the real part of the variable $U$, while its
imaginary part is kept unfixed.
Due to this,
the value ${\rm Re} ({\cal C} {\cal G}_0)$ is unfixed.
However, this should be very large under the supergravity approximation.
The only one condition is that ${\rm Re} ({\cal C} {\cal G}_0)$
does not vanish in order to realize a well-defined K\"{a}hler
potential (\ref{IIAKahlerN=1}). 
This result again differs from that of \cite{Anguelova:2008fm}.
There would be at least two possibilities:
(i) The prepotential ${\cal F}$ in (\ref{IIA-F3}) would not be an
appropriate form to find a Minkowski vacuum.
(ii) The attractor equations \cite{Anguelova:2008fm} based on the work 
\cite{Kallosh:2005ax} might not be the
most generic equations to find all flux vacua.
It would be interesting to fill gaps between our result and that of
\cite{Anguelova:2008fm}.


\section{Example 3: models on $SU(3)$-structure manifold without Ramond-Ramond flux charges}
\label{section-RR0-nongeom0}

Here let us analyze a model compactified without the Ramond-Ramond
flux charges and nongeometric flux charges.
In this model we set 
\begin{align}
e_{\text{RR} \check{A}} \ &= \ 0
\, , \ls
m_{\text{RR}}^{\check{A}} \ = \ 0
\, , \ls
p_0{}^{\check{A}} \ = \ 0
\, , \ls
q^{0 \check{A}} \ = \ 0 
\, .
\end{align}
The total superpotential ${\cal W}$ and its covariant
derivatives are reduced to 
\bsubeq \label{single-RR0-nongeom0-WQ}
\begin{align}
{\cal W} \ &= \ U {\cal W}^{\cal Q}
\, , \\
{\cal W}^{\cal Q}
\ &= \ 
- e_{00} - e_{0 \check{a}} t^{\check{a}}
\, , \\
D_{\check{a}} {\cal W}
\ &= \ 
U D_{\check{a}} {\cal W}^{\cal Q}
\, , \ls
D_U {\cal W}
\ = \ 
i \, \frac{{\rm Re} U}{{\rm Im} U}
{\cal W}^{\cal Q}
\, , \\
D_{\check{b}} D_{\check{a}} {\cal W}
\ &= \ 
i \, U \, C_{\check{a} \check{b} \check{c}} (K_+)^{\check{c} \ol{\check{d}}}
D_{\ol{\check{d}}} \ol{\cal W}{}^{\cal Q}
\, , \\
D_U D_{\check{a}} {\cal W}
\ &= \ 
i \, \frac{{\rm Re} U}{{\rm Im} U}
D_{\check{a}} {\cal W}^{\cal Q}
\, , \ls
D_U D_U {\cal W}
\ = \ 
- \frac{U + 2 \ol{U}}{2({\rm Im} U)^2}
{\cal W}^{\cal Q}
\, .
\end{align}
\esubeq 
Let us first 
consider the case that $D_P {\cal W} = 0$ is satisfied.
Next we try to find a possibility that a consistent
non-supersymmetric solution which satisfies 
$D_P {\cal W} \neq 0$ with $\del_P V = 0$. 
Actually we find later that there are neither supersymmetric nor
non-supersymmetric solutions.


\subsection{Supersymmetric vacua}

In a supersymmetric solution, the equations 
$D_{\check{a}} {\cal W} = 0$ and $D_U {\cal W} = 0$ are satisfied. 
We again impose ${\rm Im} \, U \neq 0$ to find a solution with finite
curvature.
Actually this configuration is analogous to the case in heterotic string
theory compactifications in the presence of $H$-flux\footnote{Precisely speaking,
  the condition $\d H \neq 0$ is necessary to see a supersymmetric
  flux vacua in heterotic theory \cite{Kimura:2006af}.}.


For simplicity, let us first consider a single modulus model $t^{\check{a}} \equiv t$.
In this case the covariant derivative is reduced to
\begin{align}
D_t {\cal W}^{\cal Q}
\ &= \ 
\frac{1}{t - \ol{t}} \Big(
e_0 (2 t + \ol{t}) + 3 e_{00}
\Big)
\, .
\end{align}
Then we find 
\begin{align}
2 t + \ol{t} \ &= \ 
- \frac{3 e_{00}}{e_0}
\, ,
\end{align}
where the right-hand side is a real value.
This implies the solution $t$ should be real, while this is inadmissible
because the curvature (\ref{curve-WR-real-singular}) becomes singular at that point.
Thus we find there are no consistent supersymmetric solutions
which satisfy $D_t {\cal W}^{\cal Q} = 0$.
In the same way, we also find that there are no consistent solutions
of ${\cal W}^{\cal Q} = 0$ because
${\cal W}^{\cal Q} = - ( e_{00} + e_{0} t_1 ) - i e_{0} t_2$
can be zero if and only if $t_2 = 0$, which gives rise to singular curvature.
Then we conclude that there are no supersymmetric solutions in 
the single modulus model.


Next we study so-called the $stu$-model
given by the three local variables:
\begin{align}
{\cal F} \ &= \ \frac{X^s X^t X^u}{X^0}
\, , \ls
X^s \ = \ X^0 s 
\, , \ls
X^t \ = \ X^0 t 
\, , \ls
X^u \ = \ X^0 u 
\, .
\end{align}
We set $X^0 = 1$.
The superpotential ${\cal W}^{\cal Q}$, the
K\"{a}hler potential $K_+$, and other functions are described as 
\bsubeq \label{STU-fns}
\begin{align}
{\cal W}^{\cal Q}
\ &= \ 
- e_{00} - e_{0s} s - e_{0t} t - e_{0u} u
\, , \\
K_+ \ &= \ 
- \log \big( - i (s - \ol{s}) (t - \ol{t}) (u - \ol{u}) \big)
\, , \\
\del_s K_+ \ &= \ 
- \frac{1}{s - \ol{s}}
\, , \ls
\del_t K_+ \ = \ 
- \frac{1}{t - \ol{t}}
\, , \ls
\del_u K_+ \ = \ 
- \frac{1}{u - \ol{u}}
\, , \\
(K_+)^{\check{a} \ol{\check{b}}}
\ &= \ 
- \text{diag.} \big( (s - \ol{s})^2 , (t - \ol{t})^2 , (u - \ol{u})^2 \big)
\, , \\
C_{stu} \ &= \ 
\frac{i}{(s - \ol{s}) (t - \ol{t}) (u - \ol{u})}
\, , \\
D_s {\cal W}
\ &= \ 
\frac{U}{s - \ol{s}} \Big(
e_{00} + e_{0s} \ol{s} + e_{0t} t + e_{0u} u 
\Big)
\, , \ls
D_t {\cal W}
\ = \ 
\frac{U}{t - \ol{t}} \Big(
e_{00} + e_{0s} s + e_{0t} \ol{t} + e_{0u} u 
\Big)
\, , \\
D_u {\cal W}
\ &= \ 
\frac{U}{u - \ol{u}} \Big(
e_{00} + e_{0s} s + e_{0t} t + e_{0u} \ol{u} 
\Big)
\, , \\
D_s D_t {\cal W}
\ &= \ 
\frac{u - \ol{u}}{(s - \ol{s}) (t - \ol{t})} U 
D_{\ol{u}} \ol{\cal W}^{\cal Q}
\, , \ls
D_t D_u {\cal W}
\ = \ 
\frac{s - \ol{s}}{(t - \ol{t}) (u - \ol{u})} U 
D_{\ol{s}} \ol{\cal W}^{\cal Q}
\, , \\
D_u D_s {\cal W}
\ &= \ 
\frac{t - \ol{t}}{(u - \ol{u})(s - \ol{s})}
D_{\ol{t}} \ol{\cal W}^{\cal Q}
\, .
\end{align}
\esubeq
Expanding $s = s_1 + i s_2$, $t = t_1 + i t_2$ and $u = u_1 + i u_2$,
we rewrite the supersymmetry conditions:
\bsubeq
\begin{alignat}{3}
0 \ &= \ D_s {\cal W}^{\cal Q}
& \ls &\to & \ls &\left\{
\renewcommand{\arraystretch}{1.4}
\begin{array}{l}
\dps 0 \ = \ \frac{1}{2 s_2} \Big( -e _{0s} s_2 + e_{0t} t_2 + e_{0u} u_2 \Big)
\\
\dps 0 \ = \ - \frac{1}{2 s_2} \Big(
e_{00} + e_{0s} s_1 + e_{0t} t_1 + e_{0u} u_1 
\Big)
\end{array} \right.
\\
0 \ &= \ D_t {\cal W}^{\cal Q}
& \ls &\to & \ls &\left\{
\renewcommand{\arraystretch}{1.4}
\begin{array}{l}
\dps 0 \ = \ \frac{1}{2 t_2} \Big( e _{0s} s_2 - e_{0t} t_2 + e_{0u} u_2 \Big)
\\
\dps 0 \ = \ - \frac{1}{2 t_2} \Big(
e_{00} + e_{0s} s_1 + e_{0t} t_1 + e_{0u} u_1 
\Big)
\end{array} \right.
\\
0 \ &= \ D_u {\cal W}^{\cal Q}
& \ls &\to & \ls &\left\{
\renewcommand{\arraystretch}{1.4}
\begin{array}{l}
\dps 0 \ = \ \frac{1}{2 u_2} \Big( e _{0s} s_2 + e_{0t} t_2 - e_{0u} u_2 \Big)
\\
\dps 0 \ = \ - \frac{1}{2 u_2} \Big(
e_{00} + e_{0s} s_1 + e_{0t} t_1 + e_{0u} u_1 
\Big)
\end{array} \right.
\end{alignat}
\esubeq
The solution is given by
\begin{align}
- e_{0s} s_1 \ &= \ 
e_{00} + e_{0t} t_1 + e_{0u} u_1 
\, , \ls
t_1 , u_1 : \ \ \text{unfixed}
\, , \ls
e_{0s} s_2 \ = \ 
e_{0t} t_2 \ = \ 
e_{0u} u_2 \ = \ 0
\, .
\end{align}
In order to obtain the finite curvature, we should impose 
$s_2 \neq 0$, $t_2 \neq 0$ and $u_2 \neq 0$. 
This implies $e_{0s} = e_{0t} = e_{0u} = 0$ and then $e_{00} = 0$.
This solution is interpreted as a Calabi-Yau three-fold 
in the absence of fluxes.
In such a configuration 
the superpotential ${\cal W}^{\cal Q}$ becomes trivial.
Thus we conclude that there are no non-trivial solutions to realize 
supersymmetric flux vacua in the $stu$-model.

Even though we increase the number of moduli fields $t^{\check{a}}$,
we cannot find any consistent solutions to
realize supersymmetric flux vacua with the finite curvature as far as we
restrict the prepotential in the form as (\ref{IIA-F3}).
Then we have to modify the form (\ref{IIA-F3}). 
This will be discussed in the next section.


\subsection{Non-supersymmetric vacua}

Here we search a non-supersymmetric solution.
In this case we have to solve the differential equation 
$\del_P V_{\cal W} = 0$ itself.


Let us again consider the single modulus model.
In this case the functions have already been given
in (\ref{IIA-single-fns}).
The first derivatives of the scalar potential $\del_P V_{\cal W}$
(\ref{1deriv-VW}) are 
\bsubeq \label{single-nonSUSY-deriv-V}
\begin{align}
\e^{- K} \del_t V_{\cal W}
\ &= \ 
\frac{2 e_0 |U|^2}{3} \Big( e_0 (t + 2 \ol{t}) + 3 e_{00} \Big)
+ \frac{2 ({\rm Im} U)^2}{t - \ol{t}}
\Big( e_{00} + e_0 \ol{t} \Big)
\Big( e_0 (2 t + \ol{t}) + 3 e_{00} \Big) 
\, , \\
\e^{- K} \del_U V_{\cal W}
\ &= \ 
- {\rm Re} U \left( 
1 + i \frac{{\rm Re} U}{{\rm Im} U}
\right) 
\left(
- \frac{1}{3} \big| e_0 (2 t + \ol{t}) + 3 e_{00} \big|^2
+ \big| e_{00} + e_0 t \big|^2
\right)
\, .
\end{align}
\esubeq
These two complex equations give four real equations whose solutions are
\begin{align}
{\rm Re} U \ &= \ 0
\, , \ls
{\rm Im} U, \, t_1 : \ \ \text{unfixed}
\, , \ls
t_2 \ = \ 
\pm \frac{3 (e_{00} + e_0 t_1)}{e_0} 
\sqrt{-\frac{1}{5}}
\, .
\end{align}
This is inconsistent with $t_2 \in {\mathbb R}$.
Then we conclude that there are no consistent solutions which satisfy $\del_P
V_{\cal W} = 0$ in the search of non-supersymmetric vacua in the
single modulus model.


Next we consider the $stu$-model with functions (\ref{STU-fns}).
The derivatives of the scalar potential are
\bsubeq
\begin{align}
\e^{- K} \del_s V_{\cal W}
\ &= \ 
- \frac{2 |U|^2}{s - \ol{s}}
\Big(
e_{00} + e_{0s} \ol{s} + e_{0t} t + e_{0u} \ol{u}
\Big)
\Big(
e_{00} + e_{0s} \ol{s} + e_{0t} \ol{t} + e_{0u} u
\Big)
\nn \\
\ & \ \ \ \ 
+ \frac{2 |U|^2}{s - \ol{s}}
\Big(
e_{00} + e_{0s} \ol{s} + e_{0t} t + e_{0u} u
\Big)
\Big(
e_{00} + e_{0s} s + e_{0t} \ol{t} + e_{0u} \ol{u}
\Big)
\nn \\
\ & \ \ \ \ 
- \frac{(U - \ol{U})^2}{2 (s - \ol{s})}
\Big(
e_{00} + e_{0s} \ol{s} + e_{0t} \ol{t} + e_{0u} \ol{u}
\Big)
\Big(
e_{00} + e_{0s} \ol{s} + e_{0t} t + e_{0u} u
\Big)
\, , \\
\e^{- K} \del_t V_{\cal W}
\ &= \ 
- \frac{2 |U|^2}{t - \ol{t}}
\Big(
e_{00} + e_{0s} s + e_{0t} \ol{t} + e_{0u} \ol{u}
\Big)
\Big(
e_{00} + e_{0s} \ol{s} + e_{0t} \ol{t} + e_{0u} u
\Big)
\nn \\
\ & \ \ \ \ 
+ \frac{2 |U|^2}{t - \ol{t}}
\Big(
e_{00} + e_{0s} s + e_{0t} \ol{t} + e_{0u} u
\Big)
\Big(
e_{00} + e_{0s} \ol{s} + e_{0t} t + e_{0u} \ol{u}
\Big)
\nn \\
\ & \ \ \ \ 
- \frac{(U - \ol{U})^2}{2 (t - \ol{t})}
\Big(
e_{00} + e_{0s} \ol{s} + e_{0t} \ol{t} + e_{0u} \ol{u}
\Big)
\Big(
e_{00} + e_{0s} s + e_{0t} \ol{t} + e_{0u} u
\Big)
\, , \\
\e^{- K} \del_u V_{\cal W}
\ &= \ 
- \frac{2 |U|^2}{u - \ol{u}}
\Big(
e_{00} + e_{0s} \ol{s} + e_{0t} t + e_{0u} \ol{u}
\Big)
\Big(
e_{00} + e_{0s} s + e_{0t} \ol{t} + e_{0u} \ol{u}
\Big)
\nn \\
\ & \ \ \ \ 
+ \frac{2 |U|^2}{u - \ol{u}}
\Big(
e_{00} + e_{0s} s + e_{0t} t + e_{0u} \ol{u}
\Big)
\Big(
e_{00} + e_{0s} \ol{s} + e_{0t} \ol{t} + e_{0u} u
\Big)
\nn \\
\ & \ \ \ \ 
- \frac{(U - \ol{U})^2}{2 (u - \ol{u})}
\Big(
e_{00} + e_{0s} \ol{s} + e_{0t} \ol{t} + e_{0u} \ol{u}
\Big)
\Big(
e_{00} + e_{0s} s + e_{0t} t + e_{0u} \ol{u}
\Big)
\, , \\
\e^{- K} \del_U V_{\cal W}
\ &= \ 
- \frac{\ol{U} (U + \ol{U})}{U - \ol{U}}
\Big( 
e_{00} + e_{0s} \ol{s} + e_{0t} t + e_{0u} u
\Big)
\Big( 
e_{00} + e_{0s} s + e_{0t} \ol{t} + e_{0u} \ol{u}
\Big)
\nn \\
\ & \ \ \ \ 
- \frac{\ol{U} (U + \ol{U})}{U - \ol{U}}
\Big( 
e_{00} + e_{0s} s + e_{0t} \ol{t} + e_{0u} u
\Big)
\Big( 
e_{00} + e_{0s} \ol{s} + e_{0t} t + e_{0u} \ol{u}
\Big)
\nn \\
\ & \ \ \ \ 
- \frac{\ol{U} (U + \ol{U})}{U - \ol{U}}
\Big( 
e_{00} + e_{0s} s + e_{0t} t + e_{0u} \ol{u}
\Big)
\Big( 
e_{00} + e_{0s} \ol{s} + e_{0t} \ol{t} + e_{0u} u
\Big)
\nn \\
\ & \ \ \ \ 
+ \frac{\ol{U} (U + \ol{U})}{U - \ol{U}}
\Big( 
e_{00} + e_{0s} s + e_{0t} t + e_{0u} u
\Big)
\Big( 
e_{00} + e_{0s} \ol{s} + e_{0t} \ol{t} + e_{0u} \ol{u}
\Big)
\, .
\end{align}
\esubeq
All these equations should vanish to realize a non-supersymmetric solution.
Computing them,
we obtain two solutions:
\bsubeq
\begin{gather}
\left\{
\begin{array}{c}
\dps {\rm Re} U \ = \ 0
\, , \ \ \ 
{\rm Im} U , s_1, t_1, u_1: \ \ \text{unfixed}
\, , \ \ \ 
- {\rm Re} ({\cal W}^{\cal Q})
\ \equiv \ 
e_{00} + e_{0s} s_1 + e_{0t} t_1 + e_{0u} u_1
\, , \\
\dps 
s_2 \ = \ 
- \frac{{\rm Re} ({\cal W}^{\cal Q})}{\sqrt{-5} e_{0s}}
\, , \ls
t_2 \ = \ 
- \frac{{\rm Re} ({\cal W}^{\cal Q})}{\sqrt{-5} e_{0t}}
\, , \ls
u_2 \ = \ 
- \frac{{\rm Re} ({\cal W}^{\cal Q})}{\sqrt{-5} e_{0u}}
\end{array}
\right\}
\\
\left\{
\begin{array}{c}
\dps {\rm Im} U , u_1, t_1, u_1, u_2 : \ \ \text{unfixed}
\, , \ls 
\omega \ \equiv \ - \frac{1 - \sqrt{3} i}{2}
\, , \\
\dps 
{\rm Re} U \ = \ i {\rm Im} U
\, , \ \ \ 
s_1 \ = \ 
- \frac{e_{00} + e_{0t} t_1 + e_{0u} u_1}{e_{0s}}
\, , \ \ \ 
s_2 \ = \ - \frac{e_{0u} u_2}{e_{0s}} \omega^2
\, , \ \ \ 
t_2 \ = \ 
\frac{e_{0u} u_2}{e_{0t}} \omega
\end{array}
\right\}
\end{gather}
\esubeq
Both of them are inconsistent. We conclude that there are
no consistent solutions to realize non-supersymmetric flux vacua in the $stu$-model.
In principle, the structures of the equations in multi moduli models
  are same as the $stu$-model.
Then we also find that there are no non-supersymmetric flux vacua in a
generic multi moduli model.

We summarize that there are no consistent solutions to realize
four-dimensional spacetime vacua only in the presence of geometric
fluxes if the prepotential is restricted to (\ref{IIA-F3}).
It is inevitable to introduce corrections to the prepotential ${\cal F}$.


\section{Example 4: another model on $SU(3)$-structure manifold}
\label{section-RR0-nongeom0-deform}

Since we could not find any consistent solutions 
in section \ref{section-RR0-nongeom0}, 
we have to introduce a deformation in
the prepotential ${\cal F}$ in the following way\footnote{Insertion of
the corrections is also discussed in \cite{Palti:2007pm}.}:
\begin{align}
{\cal F} (X) \ &= \ 
D_{\check{a} \check{b} \check{c}} \frac{X^{\check{a}} X^{\check{b}} X^{\check{c}}}{X^0}
+ \wt{\cal F} (X)
\, , \label{prepotential-deform1}
\end{align}
where 
$\wt{\cal F} (X)$ is also 
a holomorphic function of the projective coordinates of degree two.
Here we focus on a single modulus model $t^{\check{a}} \equiv t$.
The prepotential ${\cal F}$ and the K\"{a}hler potential are given by
\bsubeq
\begin{align}
{\cal F} \ &= \ \frac{X^t X^t X^t}{X^0} +  \wt{\cal F} (X)
\, , \ls
X^t \ = \ X^0 t
\, , \\
K_+ \ &= \ 
- \log \big( - i (t - \ol{t}{}^3) + iN \big)
\, , \\
N \ &\equiv \ 
\wt{\cal F}_0 - \ol{\wt{\cal F}}_0 + \ol{t} \wt{\cal F}_t - t \ol{\wt{\cal F}}_t
\, , \ls
\del_t N \ \equiv \ 
\wt{\cal F}_{0t} + \ol{t} \wt{\cal F}_{tt} - \ol{\wt{\cal F}}_t
\, .
\end{align}
\esubeq
For a minimal setup
we introduce the deformed term $\wt{\cal F}$ in the following form:
\begin{align}
\wt{\cal F} \ &= \ N_1 \frac{(X^t)^4}{(X^0)^2}
\, . \label{F-correction_n1}
\end{align}
Then the function $N$ and its derivatives are 
\bsubeq
\begin{align}
N \ &= \ 
-2 \Big( N_1 t^4 - \ol{N}_1 \ol{t}{}^4
- 2 N_1 t^3 \ol{t}
+ 2 \ol{N}_1 t \ol{t}{}^3
\Big)
\, , \\
\del_t N
\ &= \ 
- 4 \Big( 2 N_1 t^3 
- 3 N_1 t^2 \ol{t}
+ \ol{N}_1 \ol{t}{}^3
\Big)
\, .
\end{align}
\esubeq
The function $N$ gives a consistent solution of $D_t {\cal W}^{\cal Q} = 0$:
\bsubeq \label{SUSY_sol_deform}
\begin{align}
t_{1*} \ &= \ 
- \frac{2 e_{00}}{e_0}
\, , \ls t_{2*} \ = \ 0 
\, , \ls
{\rm Re} U_* \ = \ 0
\, , \\
{\cal W}_*^{\cal Q}
\ &= \ 
e_{00}
\, , \\
M_* \ &= \ 
- \frac{64 i (e_{00})^4 {\rm Im} N_1}{(e_0)^4}
\, , \ls
(K_+)_* 
\ = \ 
- \log \Big( 
- \frac{64 (e_{00})^4 {\rm Im} N_1}{(e_0)^4}
\Big)
\, , \\
R^t{}_{tt\ol{t}} \big|_*
\ &= \ 
\frac{(e_0)^2}{256} 
\frac{832 (e_{00})^2 ({\rm Im} N_1)^2 
- 144 e_{00} e_0 {\rm Re} N_1 
+ 576 (e_{00})^2 ({\rm Re} N_1)^2
+ 9 (e_{0})^2}{(e_{00})^4 {\rm Im} N_1}
\, .
\end{align}
\esubeq
This is indeed a solution which gives the finite curvature.
We have to set ${\rm Im} N_1$ to be negative definite, 
otherwise the K\"{a}hler potential $K_+$ is ill-defined.
The scalar potential is evaluated: 
\begin{align}
V_* \ &= \ 
- 3 \, \e^{K} |{\cal W}_*|^2
\ = \ 
\frac{1}{[ {\rm Re} ({\cal C} {\cal G}_0)]^2}
\frac{3(e_0)^4}{16 (e_{00})^2 {\rm Im} N_1}
\, . \label{AdS_deform-singleV}
\end{align}
Due to the condition ${\rm Im} N_1 < 0$, 
$V_*$ provides the negative cosmological constant.
In order to satisfy the supergravity approximation,
the value ${\rm Re} ({\cal C} {\cal G}_0)$ should be very large.
This is nothing but the solution to realize a supersymmetric AdS
vacuum in the compactification on the $SU(3)$-structure manifold.

We also find that this scalar potential and the curvature tensor 
go to infinity when we take the
limit $N_{1} \to 0$.
The geometric flux charges deform the internal space.
This is the reason why we could not find any solutions in section 
\ref{section-RR0-nongeom0}.
This result again differs from that of \cite{Anguelova:2008fm}.
If we set the torsion charge $e_0$ to be zero, the internal manifold is
reduced to a Calabi-Yau three-fold with $H$-flux charges
$e_{00}$. Here we cannot take the large volume limit 
(\ref{IIA-F3}) caused by the existence of $H$-flux. 
Then the deformation (\ref{prepotential-deform1}) is inevitable.
In this case the cosmological constant vanishes and a
Minkowski vacuum appears. 
This is consistent with that of \cite{Anguelova:2008fm}.


\section{Summary and discussions}
\label{Discussions}

In this paper we studied supersymmetric vacua in four-dimensional
${\cal N}=1$ supergravity derived from type IIA string theory
compactified on generalized geometries with $SU(3) \times SU(3)$ structures. 
We started with a generic form of the 
scalar potential in ${\cal N}=1$ supergravity 
which contains a superpotential and D-terms.
The superpotential is built from two parts; 
one is given by Ramond-Ramond flux charges, 
the other by (non)geometric flux charges. 
We referred to the former as the Ramond-Ramond flux superpotential,
and to the latter as the (non)geometric flux superpotential.

To make the discussion clear,
we first addressed a simple model with
a prepotential given by the intersection number 
in a way analogous to a model derived from a 
compactification on Calabi-Yau three-fold in the large volume limit. 
We obtained two supersymmetric vacua characterized by
discriminants of the superpotentials.
If the discriminants of the Ramond-Ramond flux
superpotential and of the {(non)geometric flux superpotential} 
are positive, a supersymmetric AdS vacuum is realized. 
The cosmological
constant is given by the square root of the discriminant of the superpotential.
This situation is akin to flux vacua attractors in type IIB theory.
On the other hand, if both of these two discriminants are negative,
the cosmological constant vanishes and a supersymmetric Minkowski
vacuum appears. 

Next we explored consistent supersymmetric vacua in the absence
of Ramond-Ramond flux charges. 
In a simple model on generalized geometry with
$SU(3) \times SU(3)$ structures, 
we again obtained a supersymmetric AdS
vacuum with a negative cosmological constant.
If the nongeometric flux charges are turned off in a situation
where the prepotential is described only in terms of the intersection number, 
there exist neither supersymmetric nor non-supersymmetric solutions.
Then we analyzed another model which has a prepotential with a
deformation term, obtaining a consistent supersymmetric AdS vacuum.
This implies that 
a model compactified on an $SU(3)$-structure manifold with torsion 
in the absence of Ramond-Ramond flux charges
differs from a model given by Calabi-Yau compactification 
in the large volume limit.

There are four interesting issues 
which deserve further study in flux compactification scenarios on
generalized geometries:
(i) In this paper there is no way to fix 
the real part of the modulus $U$ in the supersymmetric AdS vacua,
partly because 
we restricted the number of complex variables $U^{\check{I}}$ to one.
If one incorporates more than one variable,
there might appear a richer structure in various functions,
especially in the second derivatives of the K\"{a}hler potential.
In addition, it is also worth considering non-perturbative corrections to stabilize
all moduli.
(ii) In a generic configuration with Ramond-Ramond
flux charges and nongeometric flux charges,
we restricted the form of the prepotential governing the chiral
scalar variables $t^{\check{a}}$ in the same way as one does for the Calabi-Yau
compactification in the large volume limit.
This corresponds to a model compactified on a parallelizable
twisted torus. 
One should also consider models arising from more generic
prepotentials to understand lower-dimensional
effective theories of compactifications on (non)geometric string backgrounds. 
(iii) The Bianchi identity of form fluxes should also be considered
seriously to study consistent configurations of D-branes and orientifold planes
wrapped on the internal space \cite{Grana:2006kf}.
(iv) Duality transformations on generalized geometries
are crucial in elucidating the stringy origin of nongeometric fluxes in a
  more explicit way \cite{Grana:2008yw}. 
This way also be helped by use of doubled space formalism
  \cite{Hull:2004in, Hull:2005hk, Shelton:2005cf, Hull:2007jy, Dall'Agata:2007sr}. 
Duality transformations and nongeometric compactifications may also
ultimately lead to a complete classification of lower-dimensional
gauged supergravities which are not derived from higher-dimensional
supergravities compactified on conventional geometries.


\section*{Acknowledgements}

The author would like to thank Tohru Eguchi for 
illuminating discussions in the early stages of this work.


\begin{appendix}

\section*{Appendix}


\section{Supersymmetry parameters}
\label{SUSYab}

We consider type IIA string theory compactified on
generalized geometries.
Let us assume that ten-dimensional metric is given by
$\d s_{10}^2 = \e^{2 A} \, g_{\mu \nu} \, \d x^{\mu} \d x^{\nu} 
+ g_{mn} \, \d y^m \d y^n$,
where $g_{\mu \nu}$ and $g_{mn}$ are the metric of the four-dimensional
spacetime $\mathscr{M}_{3,1}$ and that of the six-dimensional space
$\mathscr{M}$, respectively. 
We also introduced a warp factor $A$.
For simplicity, the warp factor is a constant.
The ten-dimensional supersymmetry parameters $\epsilon^1$ and 
$\epsilon^2$ are split into two parts:
\begin{align}
\epsilon^1 \ = \ \ve_1 \otimes \ol{a} \eta_-^1 + \ve^1 \otimes a \eta_+^1
\, \ls 
\epsilon^2 \ = \ \ve_2 \otimes b \eta_+^2 + \ve^2 \otimes \ol{b} \eta_-^2
\, . \label{SUSY-decompose}
\end{align}
Here $\ve_{\cal A}$ with indices
${\cal A} = 1,2$ are Weyl fermions as 
the four-dimensional supersymmetry parameters whose charge conjugates are 
$\ve_{\cal A}^c \equiv \ve^{\cal A}$.
The $\eta_{\pm}^{\cal A}$ are $SU(4)$ Weyl spinors in
the six-dimensional internal space with 
$(\eta_{\pm}^{\cal A})^c = (\eta_{\pm}^{\cal A})^*$.
The chirality of $\epsilon^1$ ($\epsilon^2$) in type IIA theory
is negative (positive) \cite{Grana:2006hr, Cassani:2007pq}.
The two complex scale parameters $a$ and $b$ are
normalization factors \cite{Grana:2006kf, Cassani:2007pq} with
$|a|^2 + |b|^2 = c_+$ and $|a|^2 - |b|^2 = c_-$.
Without loss of generality we can set $c_+ = 1$.
Indeed, the coefficients $a$ and $b$ 
would be related to the warp factor $A$ in ${\cal N}=1$ vacua \cite{Grana:2005sn}. 
In order to obtain 
${\cal N}=2$ and ${\cal N}=1$ supersymmetries
in four-dimensional spacetime, 
the $SU(4)$ spinors are reduced to $SU(3)$ invariant spinors
which are interpreted as Killing spinors on $\mathscr{M}$.


\section{Generalized geometries with $SU(3) \times SU(3)$ structures}

\subsection{Generalized complex structures and pure spinors}

In the splitting of type IIA supersymmetry parameters
(\ref{SUSY-decompose}), there emerges a pair of $SU(3)$ invariant
spinors $\eta_+^1$ and $\eta_+^2$. 
These two spinors are related to each other via the expression \cite{Grana:2006hr}
\begin{align}
\eta_+^2 \ &= \ 
c_{\parallel} \eta_+^1 + c_{\perp} (v + i v')^m \gamma_m \eta_-^1
\, , \ls
|c_{\parallel}|^2 + |c_{\perp}|^2 = 1
\, , \label{eta_12}
\end{align}
where $\gamma_{m}$ is the ${\rm Cliff} (6)$ gamma matrix acting on
$\eta^{\cal A}_{\pm}$.
The two vectors $v$ and $v'$ are defined by the bilinear form of the
spinors as
$(v - i v')^m = \eta_+^{1 \dagger} \gamma^m \eta_-^2$.
The coefficients $c_{\parallel}$ and $c_{\perp}$ depend on the coordinates of
the internal space $\mathscr{M}$.
This pair of spinors defines a pair of
$SU(3)$-structure groups, where the structure group is the group 
in which the transition functions of the tangent bundle $T\mathscr{M}$
take their values. 
If $c_{\perp} = 0$ at any point on $\mathscr{M}$, the two spinors
coincide with each other and the structure group is reduced to a single $SU(3)$.
As usual one defines the almost complex structures 
in terms of the $SU(3)$ invariant spinors as
$(J^{\cal A})^m{}_n = - 2i \, \eta_+^{{\cal A}\dagger} \gamma^m{}_n
\eta_+^{\cal A}$.
If $\eta_+^1 = \eta_+^2$ at any points, the almost complex structures
$J^1$ also coincides with $J^2$. 
We refer to $\mathscr{M}$ with a single almost complex structure as a
manifold with a single $SU(3)$-structure, or simply an $SU(3)$-structure manifold.
On the other hand, if $c_{\perp} \neq 0$ at some points on $\mathscr{M}$,
there exists a pair of almost complex structures on $\mathscr{M}$, and we refer 
to this as a manifold with a pair of $SU(3)$-structures.

To go beyond an ordinary almost complex structure,
one considers a space $T\mathscr{M} \oplus T^* \mathscr{M}$ 
and introduces generalized almost complex structures
${\cal J}_{\pm}$ which give rise to a mapping
${\cal J}_{\pm}: T\mathscr{M} \oplus T^* \mathscr{M} \to 
T\mathscr{M} \oplus T^* \mathscr{M}$.
Since the basis of the space $T\mathscr{M} \oplus T^* \mathscr{M}$ is 
given by $\{ \d x^m \w , \iota_{\del_n} \}$, the signature of this
space is $(6,6)$.
Let us first describe ${\cal J}_{\pm}$ by means of 
sections of spinor bundles 
associated with $T\mathscr{M} \oplus T^* \mathscr{M}$:
\begin{align}
{\cal J}_{\pm}^{\Lambda}{}_{\Sigma} 
\ &= \ 
\Vev{{\rm Re} \Phi_{\pm} , \Gamma^{\Lambda}{}_{\Sigma} {\rm Re} \Phi_{\pm}}
\, . \label{GACS}
\end{align}
Here we introduced complex $SU(3,3)$ invariant spinors $\Phi_{\pm}$, 
the Cliff$(6,6)$ gamma matrix $\Gamma^{\Lambda}$ and its
antisymmetrized product $\Gamma^{\Lambda \Sigma} = \half
(\Gamma^{\Lambda} \Gamma^{\Sigma} - \Gamma^{\Sigma} \Gamma^{\Lambda})$,
where the indices are raised and lowered with the $SO(6,6)$ invariant
metric $L^{\Lambda \Sigma}$.
Since the irreducible representation of $Spin(6,6)$ spinor is
Majorana-Weyl, $\Phi_+$ ($\Phi_-$) can be assigned to a Weyl
spinor with positive (negative) chirality.
The two Weyl spinor bundles on 
$T\mathscr{M} \oplus T^* \mathscr{M}$ 
are {\it isomorphic} to the spaces of 
even/odd forms $\w^{\text{even/odd}} T^* \mathscr{M}$.
The $SU(3,3)$ invariant Weyl spinors $\Phi_{\pm}$ are
pure since they are
annihilated by half of the ${\rm Cliff}(6,6)$ gamma
matrices $\Gamma_{\Lambda}$. 
Due to the isomorphism, 
the bracket in (\ref{GACS}) can be expressed by the Mukai pairing
\begin{align}
\Vev{A_p, B_q}
\ \equiv \ 
[A_p \w \lambda (B_q)]_{\text{top form}}
\, , \ls
\lambda (B_q)  \ \equiv \
(-1)^{[\frac{q}{2}]} B_q
\, , \label{Mukai}
\end{align}
where $A_p$ and $B_q$ are arbitrary $p$-form and $q$-form, respectively.
When a generalized almost complex structure ${\cal J}$ is defined, 
we refer to the space $\mathscr{M}$ as a generalized (almost complex) geometry.
The two Weyl spinors $\Phi_{\pm}$ can be described in terms of 
the supersymmetry parameters $\eta_{\pm}^{\cal A}$ in (\ref{SUSY-decompose})
\cite{Grana:2006hr}:
\begin{gather}
\Phi_{\pm} \ = \ \e^{-B} \Phi^0_{\pm}
\, , \ls
\Phi^0_{\pm} \ = \ 8 \eta^1_+ \otimes \eta^{2\dagger}_{\pm}
\ \equiv \ 
\sum_{k=0}^6 \frac{1}{k!} \big( \eta_{\pm}^{2\dagger} \gamma_{m_1
  \cdots m_k} \eta_+^1 \big) \gamma^{m_k \cdots m_1}
\, ,
\end{gather}
where $B$ is a two-form. 
Actually the bilinear forms $\Phi_{\pm}^0$ satisfy the following
differential equations in ${\cal N}=1$ vacua derived from type IIA theory 
(see \cite{Grana:2005sn, Grana:2006kf, Cassani:2007pq})
\bsubeq
\begin{align}
\e^{-2 A + \phi} (\d - H \w) \big( \e^{2 A- \phi} \Phi_+^0 \big)
\ &= \ 
- 2 \ol{\mu} \, {\rm Re} \Phi_-^0
\, , \\
\e^{-2 A + \phi} (\d - H \w) \big( \e^{2 A- \phi} \ol{\Phi_-^0} \big)
\ &= \ 
- 3 i \, {\rm Im} ( \ol{\mu} \ol{\Phi_+^0})
+ \frac{1}{16} \e^{\phi} \Big[
c_- F^{\text{even}} + i c_+ * \lambda (F^{\text{even}})
\Big]
\, ,
\end{align}
\esubeq
where $F^{\text{even}} = F_0 + F_2 + F_4 + F_6$ is a sum of the Ramond-Ramond forms.
The $\mu$ gives the cosmological constant $\Lambda = - |\mu|^2$ in
four-dimensional spacetime.
Note that the structure group of the generalized geometry 
is $SU(3) \times SU(3)$ if $c_{\perp} \neq 0$ at
some points on $\mathscr{M}$, or $SU(3)$ if $c_{\perp} = 0$ at any points
on $\mathscr{M}$.

It is also known that the spaces of $\Phi_{\pm}$ are given by
special K\"{a}hler geometries
of local type \cite{Grana:2006hr}. 
This implies that the generalized geometry has 
the moduli space given by the product of the two Hodge-K\"{a}hler
geometries whose 
K\"{a}hler potentials $K_{\pm}$ are\footnote{Here these K\"{a}hler
  potentials are reduced to 
functions in four-dimensional spacetime \cite{Grana:2006hr}.}
\begin{align}
K_+ \ &= \ - \log i \int_{\mathscr{M}} \Vev{ \Phi_+ , \ol{\Phi}_+ }
\, , \ls 
K_- \ = \ - \log i \int_{\mathscr{M}} \Vev{ \Phi_- , \ol{\Phi}_- }
\, . \label{Kahler_pot_int}
\end{align}
We assign the special K\"{a}hler geometries given by $\Phi_{\pm}$ to
${\cal M}_{\pm}$, respectively.
One can introduce
projective coordinates $X^A$ and a prepotential ${\cal F}$
on ${\cal M}_+$ (and projective coordinates $Z^I$ and a prepotential
${\cal G}$ on ${\cal M}_-$).
The prepotentials ${\cal F}$ and ${\cal G}$ are functions of holomorphic
and homogeneous of degree two in the projective coordinates. 
Since the two Weyl spinors $\Phi_{\pm}$ are isomorphic to the even and
odd forms, they are expanded in terms of basis forms:
\begin{align}
\Phi_+ \ &= \ X^A \omega_A - {\cal F}_A \wt{\omega}^A
\, , \ls
\Phi_- \ = \ Z^I \alpha_I - {\cal G}_I \beta^I
\, ,
\end{align}
where $\omega_A$ and $\wt{\omega}{}^A$ are even real basis forms
(i.e., zero-, two-, four- and six-forms),
while $\alpha_I$ and $\beta^I$ are odd real basis forms 
(one-, three- and five-forms).
The coefficients are interpreted as the projective coordinates and derivatives of
the prepotentials.


\subsection{Property of Special K\"{a}hler geometries}
\label{app-HodgeKahler}

The projective coordinates and the prepotentials on the special K\"{a}hler
geometries are described in terms of period integrals of the Mukai pairing:
\bsubeq \label{purespinors}
\begin{alignat}{2}
X^A \ &= \ \int_{\mathscr{M}} \Vev{ \Phi_+ , \wt{\omega}{}^A }
\, , & \LS
{\cal F}_A \ &= \ \frac{\del {\cal F}}{\del X^A}
\ = \ 
\int_{\mathscr{M}} \Vev{ \Phi_+ , \omega_A }
\, , \\
Z^I \ &= \ \int_{\mathscr{M}} \Vev{ \Phi_- , \beta^I }
\, , & \LS
{\cal G}_I \ &= \ \frac{\del {\cal G}}{\del Z^I}
\ = \
\int_{\mathscr{M}} \Vev{ \Phi_- , \alpha_I }
\, ,
\end{alignat}
\esubeq
where we used the symplectic structure among the basis forms 
\bsubeq \label{basis-forms}
\begin{align}
\left(
\renewcommand{\arraystretch}{1.8}
\begin{array}{cc} 
\dps \int_{\mathscr{M}} \vev{\omega_A, \omega_B} &
\dps \int_{\mathscr{M}} \vev{\omega_A , \wt{\omega}{}^B} \\
\dps \int_{\mathscr{M}} \vev{\wt{\omega}{}^A, \omega_B} 
& \dps \int_{\mathscr{M}} \vev{\wt{\omega}{}^A, \wt{\omega}{}^B}
\end{array} \right)
\ &= \ 
\left(
\begin{array}{cc}
0 & \delta_A{}^B \\
- \delta^A{}_B & 0 
\end{array} \right)
\, , &
A,B \ &= \ 0,1, \dots, b^+ 
\, , \\
\left(
\renewcommand{\arraystretch}{1.8}
\begin{array}{cc}
\dps \int_{\mathscr{M}} \vev{\alpha_I , \alpha_J} 
& \dps \int_{\mathscr{M}} \vev{\alpha_I , \beta^J} \\
\dps \int_{\mathscr{M}} \vev{\beta^I , \alpha_J} 
& \dps \int_{\mathscr{M}} \vev{\beta^I , \beta^J}
\end{array} \right)
\ &= \ 
\left(
\begin{array}{cc}
0 & \delta_I{}^J \\
- \delta^I{}_J & 0 
\end{array} \right)
\, , &
I,J \ &= \ 0,1,\dots, b^-
\, .
\end{align}
\esubeq
Then the K\"{a}hler potentials $K_{\pm}$ in (\ref{Kahler_pot_int})
are described as
\begin{align}
K_+ \ &= \ - \log i \big( \ol{X}{}^A {\cal F}_A - X^A \ol{\cal F}_A \big)
\, , \ls
K_- \ = \ - \log i \big( \ol{Z}{}^I {\cal G}_I - Z^I \ol{\cal G}_I
\big)
\, .
\end{align}
We can choose a set of local coordinate frames
of ${\cal M}_{\pm}$ as
$X^A = ( X^0 , X^a ) = ( X^0, X^0 t^a )$ 
and
$Z^I = ( Z^0 , Z^i ) = ( Z^0, Z^0 z^i )$,
where $A, B, C, \dots$ and $a,b,c, \dots$ are projective and
local coordinate indices, respectively.
The properties of their functions include
\bsubeq \label{HK+}
\begin{gather}
\del_a \ \equiv \ 
\frac{\del}{\del t^a}
\, , \ls
D_c \ \equiv \ \del_c + \del_c K_+
\, , \\
{\cal F}_A \ = \ {\cal N}_{AB} X^B
\, , \ls
D_a {\cal F}_B \ = \ 
\ol{\cal N}_{BC} D_a X^C
\, , \ls
(K_+)_{a \ol{b}} \ = \ 
\del_a \del_{\ol{b}} K_+
\, , \\
\e^{K_+} (K_+)^{a \ol{b}} D_a X^C D_{\ol{b}} \ol{X}{}^D
\ = \ 
- \half [ ({\rm Im} {\cal N})^{-1} ]^{CD} 
- \e^{K_+} \ol{X}{}^C X^D
\, , \\
C_{abc}
\ = \ 
\e^{K_+} \big( \del_a X^A \big) \big( \del_b X^B \big) \big( \del_c X^C \big) 
{\cal F}_{ABC} (X)
\, , \ls
{\cal F}_{ABC} 
\ = \ 
\frac{\del^3 {\cal F}}{\del X^A \del X^B \del X^C}
\, ,
\end{gather}
\esubeq
where ${\cal N}_{AB}$ is the period matrix on the moduli space
${\cal M}_+$.
Here $D_c$ is the K\"{a}hler covariant derivative. 
Details of the special K\"{a}hler geometry can be found, for instance,
in \cite{Andrianopoli:1996cm, Grimm:2004ua, Cassani:2007pq}. 
Notice that $C_{abc}$ is a totally symmetric
K\"{a}hler covariantly holomorphic tensor on ${\cal M}_+$. 


\subsection{(Non)geometric flux charges}
\label{flux-charges}

Once the NS-NS three-form flux $H$ is incorporated into the six-dimensional internal
space $\mathscr{M}$, 
this space is no longer a Calabi-Yau three-fold\footnote{In
a very restricted case, the internal space becomes a warped Calabi-Yau manifold. Such
a geometry appears in type IIB theory flux compactification scenario
\cite{Giddings:2001yu, Grana:2005jc}.}.
Although this flux does not modify the $SU(3)$-structure group, 
a non-constant dilaton, a warp factor and torsion are induced.
We call them geometric fluxes.

In the case of the generalized geometry with $SU(3) \times SU(3)$ structures, 
we should introduce
a set of charges $p_I{}^A$ and $q^{IA}$ \cite{Grana:2006hr}, 
called the charges of ``nongeometric fluxes'' 
as well as geometric electric- and magnetic-charges $e_{IA}$ and
$m_A{}^I$ \cite{Shelton:2005cf}. 
One has to generalize the exterior derivative $\d$
to ${\cal D}$ in the following way:
\bsubeq \label{genD}
\begin{alignat}{2}
{\cal D} \omega_A \ &\sim \ m_A{}^I \alpha_I - e_{IA} \beta^I 
\, , & \LS
{\cal D} \wt{\omega}{}^A \ &\sim \ - q^{IA} \alpha_I + p_I{}^A \beta^I
\, , \ls \\
{\cal D} \alpha_I \ &\sim \ p_I{}^A \omega_A + e_{IA} \wt{\omega}{}^A
\, , & \LS
{\cal D} \beta^I \ &\sim \ q^{IA} \omega_A + m_A{}^I
\wt{\omega}{}^A
\, , 
\end{alignat}
\esubeq
where $\sim$ means equality up to terms vanishing inside the Mukai
pairing (\ref{Mukai}) in computations of the K\"{a}hler potentials and
superpotentials.
Here ${\cal D}$ is described as
${\cal D} \equiv  \d - H^{\text{fl}} \w - f \cdot - Q \cdot - R \llcorner$,
where $H^{\text{fl}}$ is the NS-NS three-form flux\footnote{The
  cohomology of the $SU(3)$-structure manifold defines the topological
  indices such as the Dirac index, the Euler characteristics and the
  Hirzebruch signature \cite{Kimura:2007xa}.}
$H^{\text{fl}} \equiv H - \d B$,
while  $f$, $Q$ and $R$ are called the (non)geometric fluxes 
acting on
an arbitrary $k$-form $C$ as
$(f \cdot C)_{m_1 \cdots m_{k+1}} \equiv
f^a{}_{[m_1 m_2} C_{|a|m_3 \cdots m_{k+1}]}$,
$(Q \cdot C)_{m_1 \cdots m_{k-1}} \equiv 
Q^{ab}{}_{[m_1} C_{|ab|m_2 \cdots m_{k-1}]}$
and
$(R \llcorner C)_{m_1 \cdots m_{k-3}} \equiv 
R^{abc} C_{abc m_1 \cdots m_{k-3}}$.
Actually the geometric flux $f$ gives a non-trivial
structure constant in gauged supergravity via the Scherk-Schwarz
compactifications \cite{Kaloper:1999yr}, while
the fluxes $Q$ and $R$ provide
the nongeometric string backgrounds \cite{Shelton:2005cf} 
via duality transformations in string theory.

Imposing the nilpotency  ${\cal D}^2 = 0$, we obtain
a set of relations among the (non)geometric flux charges:
\bsubeq \label{nongeom-Q-charges}
\begin{align}
0 \ &= \ 
q^{IA} m_A{}^J - m_A{}^I q^{IA} 
\, , &
0 \ &= \ 
p_I{}^A e_{AJ} - e_{IA} p_J{}^A
\, , &
0 \ &= \ 
p_I{}^A m_A{}^J - e_{IA} q^{JA}
\, , \\
0 \ &= \ 
q^{IA} p_I{}^B - p_I{}^A q^{IB} 
\, , &
0 &= \ 
m_A{}^I e_{IB} - e_{IA} m_B{}^I
\, , &
0 \ &= \ 
m_A{}^I p_I{}^B - e_{IA} q^{IB}
\, .
\end{align}
\esubeq


\section{Type IIA theory compactified on generalized geometry}

We analyze four-dimensional supergravity
compactified on the generalized geometry with $SU(3) \times SU(3)$ structures
by using the notation and conventions
in \cite{Cassani:2007pq}. 
First we construct ${\cal N}=1$ K\"{a}hler potential, superpotential and
D-terms in the language of ${\cal N}=2$ theory. 
Then we truncate
physical degrees of freedom via O6 orientifold projection.


${\cal N}=2$ Killing prepotentials are useful to derive the superpotential
and the D-terms.
Here we briefly review the works \cite{Grana:2006hr, Cassani:2007pq}.
The Killing prepotentials ${\cal P}^x$ appear 
in supersymmetry variations of four-dimensional
gravitinos $\psi_{{\cal A}\mu}$ as
\bsubeq \label{N2-gravitinos}
\begin{align}
\delta \psi_{{\cal A}\mu}
\ &= \ \nabla_{\mu} \ve_{\cal A} - S_{{\cal A}{\cal B}} \,
\gamma_{\mu}^{(4)} \, \ve^{\cal B} + \dots
\, , \\
S_{{\cal A}{\cal B}} \ &= \ 
\frac{i}{2} \e^{\frac{K_+}{2}} (\sigma_x)_{\cal A}{}^{\cal C} \,
\epsilon_{{\cal B}{\cal C}} \, {\cal P}^x
\ = \ 
\frac{i}{2} \e^{\frac{K_+}{2}} \left(
\begin{array}{cc}
{\cal P}^1 - i {\cal P}^2 & - {\cal P}^3 \\
- {\cal P}^3 & - {\cal P}^1 - i {\cal P}^2
\end{array} \right)
\, ,
\end{align}
\esubeq
where dots indicate irrelevant parts which do not contribute to the superpotential.
Here $\gamma^{(4)}_{\mu}$ is the Dirac gamma matrix in four dimensions,
$(\sigma_x)_{\cal A}{}^{\cal B}$ with $x = 1,2,3$ are 
the $SU(2)$ Pauli matrices, and
$\epsilon_{{\cal A}{\cal B}}$
is the $SU(2)$ invariant metric utilized to raise and lower indices ${\cal A}$.
Explicit forms of the
Killing prepotentials ${\cal P}^x$ are written
in terms of the Weyl spinors $\Phi_{\pm}$ 
and the Ramond-Ramond field strength $G$.
In the case of compactifications on the generalized geometry with
$SU(3) \times SU(3)$ structures,
these are given as follows\footnote{For detailed discussions, see
\cite{Grana:2006hr} 
for the case of generalized geometry with a single
$SU(3)$-structure, or \cite{Grana:2006hr} for that of generalized geometry
with $SU(3) \times SU(3)$ structures}:
\bsubeq \label{IIAKilling}
\begin{align}
{\cal P}^1 - i {\cal P}^2 
\ &= \ 
2 \, \e^{\frac{K_-}{2} + \varphi} 
\int_{\mathscr{M}} \Vev{ \Phi_+, {\cal D} \Phi_- }
\nn \\
\ &= \
2 \, \e^{\frac{K_-}{2} + \varphi} 
\Big[ \big( Z^I e_{IA} - {\cal G}_I m_A{}^I \big) X^A
+ \big( Z^I p_I{}^A - {\cal G}_I q^{IA} \big) {\cal F}_A \Big]
\, , \label{IIAKillingP1-P2} \\
{\cal P}^1 + i {\cal P}^2 
\ &= \
2 \, \e^{\frac{K_-}{2} + \varphi} 
\int_{\mathscr{M}} \Vev{ \Phi_+, {\cal D} \ol{\Phi}_- }
\nn \\
\ &= \  
2 \, \e^{\frac{K_-}{2} + \varphi} 
\Big[ \big( \ol{Z}{}^I e_{IA} - \ol{\cal G}_I m_A{}^I \big) X^A
+ \big( \ol{Z}{}^I p_I{}^A - \ol{\cal G}_I q^{IA} \big) {\cal F}_A \Big]
\, , \label{IIAKillingP1+P2} \\
{\cal P}^3 
\ &= \ 
- \frac{1}{\sqrt{2}} \, \e^{2 \varphi} 
\int_{\mathscr{M}} \Vev{ \Phi_+ , G }
\nn \\
\ &= \ 
\e^{2 \varphi}
\Big[ \big( e_{\text{RR}A} - \xi^I e_{IA} + \wt{\xi}_I m_A{}^I \big) X^A 
- \big( m_{\text{RR}}^A + \xi^I p_I{}^A - \wt{\xi}_I q^{IA} \big) {\cal F}_A
\Big] 
\, . \label{IIAKillingP3} 
\end{align}
\esubeq
If the six-dimensional internal space is a generalized
geometry with a single $SU(3)$-structure, the generalized
differential operator ${\cal D}$ in (\ref{IIAKilling}) 
is reduced to $\dH$.

If localized D-branes are absent,
it is convenient to define the Ramond-Ramond field strength $G$ 
as a modification of the field strength $F^{\text{even}}$ multiplied with
the exponent of the B-field \cite{Bergshoeff:2001pv, Grana:2006hr}:
\begin{gather}
F_n^{\text{even}} \ = \ (\e^B G)_n
\ = \ \d C_{n-1} - H \w C_{n-3}
\, , \ls
C \ = \ \e^B A 
\, , \ls
(\d - H \w) F^{\text{even}} \ = \ 0
\, . \label{IIA-FG-CA}
\end{gather}
In the generalized geometry with $SU(3) \times SU(3)$ structures
in the democratic description \cite{Bergshoeff:2001pv},
the Ramond-Ramond field strength $G$ 
is given in terms of the generalized differential operator ${\cal D}$ as
\begin{align}
G \ \equiv \ G_0 + G_2 + G_4 + G_6
\ &= \ G^{\text{fl}} + {\cal D} A
\, , \label{IIARR2}
\end{align}
where $G^{\text{fl}}$ and $A$ are the intrinsic part of the field
strength and the potential, respectively. Both of them are expanded
in terms of the basis of forms as\footnote{Our notation differs from that of \cite{Cassani:2007pq} by a sign, 
i.e., $e_{\text{RR} A}$ in \cite{Cassani:2007pq} becomes
$-e_{\text{RR} A}$, etc.}
\begin{align}
G^{\text{fl}} \ &= \ 
\sqrt{2} \big( m_{\text{RR}}^A \omega_A - e_{\text{RR} A}
\wt{\omega}^A \big)
\, , \ls
A \ = \ \sqrt{2} \big( \xi^I \alpha_I - \wt{\xi}_I \beta^I \big)
\, , \label{RR_A_ex}
\end{align}
where $e_{\text{RR}A}$ and $m_{\text{RR}}^A$ are electric- and 
magnetic-charges of the Ramond-Ramond fluxes, respectively.
The fields $\xi^I$ and $\wt{\xi}_I$ appear as scalar fields in four dimensions.


Let us elaborate the superpotential.
The ${\cal N}=1$ supersymmetry parameter $\ve$ is defined by the
linear combination of the two ${\cal N}=2$ supersymmetry parameters in
the following way:
\begin{align}
\ve \ &= \ \ol{n}{}^{\cal A} \ve_{\cal A}
\, . \label{N2_1_SUSY}
\end{align}
where $\ol{n}{}^{\cal A} = (a , \ol{b})$ is a two component vector
given by the coefficients $a$ and $b$ in (\ref{SUSY-decompose}).
In the same way as the linear combination (\ref{N2_1_SUSY}),
the ${\cal N}=2$ gravitinos are also linearly combined into
the ${\cal N}=1$ gravitino as
$\psi_{\mu} = \ol{n}{}^{\cal A} \psi_{{\cal A}\mu}$.
Then the ${\cal N}=1$ supersymmetry variation is described in terms of
the linear combination of the ${\cal N}=2$ variations (\ref{N2-gravitinos}) 
in such a way as
$\delta \psi_{\mu} = \nabla_{\mu} \ve - \ol{n}{}^{\cal A} S_{{\cal A}{\cal B}} \,
n^{*{\cal B}} \, \gamma^{(4)}_{\mu} \, \ve^c$.
Since this form is generically expressed as 
$\delta \psi_{\mu} = \nabla_{\mu} \ve - \e^{\frac{K}{2}} {\cal W} \, \gamma^{(4)}_{\mu} \, \ve^c$
\cite{Cassani:2007pq}, 
we obtain an explicit form of the superpotential ${\cal W}$ as
\begin{align}
\e^{\frac{K}{2}} {\cal W}
\ &= \ 
\ol{n}{}^{\cal A} S_{{\cal A}{\cal B}} \, n^{*{\cal B}}
\ = \ 
\frac{i}{2} \e^{\frac{K_+}{2}} 
\Big[
a^2 \big( {\cal P}^1 - i {\cal P}^2 \big)
- \ol{b}{}^2 \big( {\cal P}^1 + i {\cal P}^2 \big)
- 2 a \ol{b} {\cal P}^3
\Big]
\, . \label{IIA-W_proto}
\end{align}
This form, however, carries redundant information arising from 
spin 3/2 multiplets which should not appear in an ordinary 
${\cal N}=1$ supergravity. 
We define a variable $\wt{\psi}_{\mu} = b \psi_{1 \mu} - \ol{a} \psi_{2 \mu}$
which is orthogonal to the ordinary gravitino $\psi_{\mu}$ 
in order that the fermion $\wt{\psi}_{\mu}$ would be a component of the spin
3/2 multiplet. 
Imposing the invariance on the supersymmetry variation 
$\delta \wt{\psi}_{\mu} = 0$,
we obtain \cite{Cassani:2007pq}
\begin{align}
0 \ &= \ 
\e^{\frac{K_+}{2}} \Big[
a b \big( {\cal P}^1 - i {\cal P}^2 \big)
+ \ol{a} \ol{b} \big( {\cal P}^1 + i {\cal P}^2 \big)
+ c_- {\cal P}^3
\Big]
\, . \label{del_psit}
\end{align}
Substituting (\ref{del_psit}) into (\ref{IIA-W_proto}), 
we write down the correct form of the superpotential
\begin{align}
\e^{\frac{K}{2}} {\cal W}
\ &= \ 
\frac{i}{4 \ol{a} b} \e^{\frac{K_+}{2}}
\Big[
a b \big( {\cal P}^1 - i {\cal P}^2 \big)
- \ol{a} \ol{b} \big( {\cal P}^1 + i {\cal P}^2 \big)
- {\cal P}^3
\Big]
\, . \label{IIAsuper0}
\end{align}

Here we have to discuss the four-dimensional ${\cal N}=1$ K\"{a}hler
potential $K$ in the left-hand side in (\ref{IIAsuper0}).
In terms of the four-dimensional dilaton $\varphi$, 
the function $K$ is defined as 
\cite{Benmachiche:2006df}
\begin{align}
K \ &= \ K_+ + 4 \varphi
\ = \ 
- \log i \big( \ol{X}{}^A {\cal F}_A - X^A \ol{\cal F}_A \big) 
+ 4 \varphi
\, . \label{IIAKahler}
\end{align}
There is a relation among the ten-dimensional dilaton $\phi$,
the K\"{a}hler potentials $K_{\pm}$
and the four-dimensional dilaton $\varphi$ as
$\e^{- K_{\pm}} = 8 \e^{-2 \varphi + 2 \phi}$ \cite{Cassani:2007pq}.
We assumed that $\phi$ does not depend on the internal coordinates. 
Substituting (\ref{IIAKilling}) and (\ref{IIAKahler}) into (\ref{IIAsuper0}), 
we rewrite the superpotential ${\cal W}$  as\footnote{We used the same expressions of the real and the imaginary part of ${\cal C} \Phi_-$ 
as in \cite{Cassani:2007pq}.}
\bsubeq
\begin{align}
{\cal W} \ &= \ 
\frac{i}{4 \ol{a}b} \int_{\mathscr{M}} 
\Vev{ \Phi_+, \frac{1}{\sqrt{2}} G^{\text{fl}} + {\cal D} \Pi_- }
\, , \label{IIAW45} \\
\Pi_- \ &\equiv \ 
\frac{1}{\sqrt{2}} A + i \, {\rm Im} ({\cal C} \Phi_-)
\, , \ls
{\cal C} \ \equiv \ \sqrt{2} ab \, \e^{- \phi}
\ = \ 4 ab \, \e^{\frac{K_-}{2} - \varphi}
\, . \label{IIAPiC}
\end{align}
\esubeq
Here ${\cal C}$ is called a compensator of the dilaton
$\phi$ (or $\varphi$ with the K\"{a}hler potential $K_-$).
This is introduced to gauge away scale symmetry of the Weyl 
spinor $\Phi_-$ \cite{Cassani:2007pq}.
(The spaces of the spinors $\Phi_{\pm}$ are the special K\"{a}hler
  geometry of {\it local} type \cite{Grana:2006kf}.) 
Using the compensator ${\cal C}$,
we rewrite the four-dimensional dilaton $\varphi$ as
\begin{align}
\e^{- 2 \varphi}
\ &= \ \frac{|{\cal C}|^2}{16 |a|^2 |b|^2} \e^{- K_-}
\ = \ 
\frac{i}{16|a|^2|b|^2} \int_{\mathscr{M}} 
\Vev{ {\cal C} \Phi_- , \ol{\cal C} \ol{\Phi}_- }
\nn \\
\ &= \
\frac{1}{8 |a|^2|b|^2} 
\Big[ {\rm Im} ({\cal C} Z^I) {\rm Re} ({\cal C} {\cal G}_I)
- {\rm Re} ({\cal C} Z^I) {\rm Im} ({\cal C} {\cal G}_I)
\Big]
\, . \label{IIAdilaton}
\end{align}

We also rewrite the following function
in terms of the basis forms and the flux charges:
\begin{align}
\frac{1}{\sqrt{2}} G^{\text{fl}} + {\cal D} \Pi_-
\ &\sim \ 
\big( 
m_{\text{RR}}^A + U^I p_I{}^A - \wt{U}_I q^{IA}
\big) \omega_A
- \big(
e_{\text{RR}A} - U^I e_{IA} + \wt{U}_I m_A{}^I
\big) \wt{\omega}{}^A
\, .
\end{align}
It is useful to introduce 
\begin{align}
U^I \ &\equiv \ \xi^I + i \, {\rm Im} ({\cal C} Z^I)
\, , \ls
\wt{U}_I \ \equiv \ \wt{\xi}_I + i \, {\rm Im} ({\cal C} {\cal G}_I)
\, .  
\end{align}
Performing the integral in the superpotential (\ref{IIAW45}), 
we obtain the following form in the language of ${\cal N}=2$ theory:
\begin{align}
{\cal W} \ &= \ 
- \frac{i}{4 \ol{a}b} \Big[
X^A \big( e_{\text{RR} A} - U^I e_{IA} + \wt{U}_I m_A{}^I \big)
- {\cal F}_A \big( m_{\text{RR}}^A + U^I p_I{}^A - \wt{U}_I q^{IA} \big)
\Big]
\, . \label{IIAsuper2}
\end{align}
Later we truncate ${\cal N}=2$ supersymmetry and reduce physical
degrees of freedom.

In a similar way we evaluate an explicit form of the D-term from the
supersymmetry variation of the gaugino.
The supersymmetry truncation yields the ${\cal N}=1$ gaugino
$\chi^A$
as a linear combination of the ${\cal N}=2$ gauginos $\chi^{a{\cal B}}$ as
$\chi^A = 
-2 \, \e^{\frac{K_+}{2}} D_b X^A ( \ol{n}{}^{\cal A} \, 
\epsilon_{{\cal A}{\cal B}} \, \chi^{a {\cal B}} )$.
Performing ${\cal N}=1$ supersymmetry variation 
and comparing a generic form of the ${\cal N}=1$ supersymmetry
transformation rule given by
$\delta \chi^A = {\rm Im} F_{\mu \nu}^{A} \, \gamma^{\mu \nu} \ve + i D^A \ve$,
we obtain an explicit form of the D-term $D^A$ 
in the ${\cal N}=2$ language \cite{Cassani:2007pq}:
\begin{align}
D^A \ &= \ 
\e^{2 \varphi} \Big(
[ ({\rm Im} {\cal N})^{-1} ]^{AB} + 2 \, \e^{K_+} \ol{X}{}^A X^B \Big)
\nn \\
\ &\ls \times 
\Big\{
{\rm Re} ({\cal C} Z^I) [ e_{IB} + {\cal N}_{BC} p_I{}^C ]
- {\rm Re} ({\cal C} {\cal G}_I) [ m_B{}^I + {\cal N}_{BC} q^{IC} ]
\nn \\
\ &\LS
+ c_- \big[
(e_{\text{RR}B} - \xi^I e_{IB} + \wt{\xi}_I m_B{}^I)
- {\cal N}_{BC} (m_{\text{RR}}^C + \xi^I p_I{}^C - \wt{\xi}_I q^{IC})
\big]
\Big\}
\, . \label{IIAD0}
\end{align}


\section{Orientifold projection}
\label{CYorientifold}

It is necessary to introduce orientifold planes lying on the internal space
in order to realize the tadpole cancellation and to evade a no-go theorem\footnote{We do not analyze
  the Bianchi identities themselves in this paper. A detailed
  discussions can be found, for instance, in \cite{Grana:2006kf}.} 
\cite{Maldacena:2000mw}.
Due to the existence of the orientifold planes, 
the number of the supersymmetry parameters and physical
degrees of freedom are truncated.
This procedure is called the O6 orientifold projection. 
The orientifold projection affects the
coefficients $a$ and $b$ in (\ref{SUSY-decompose})
as (see, for instance, \cite{Grimm:2004ua, Grana:2006kf})
\begin{align}
a \ = \ \ol{b} \, \e^{i \theta}
\, , \ls
|a|^2 \ = \ |b|^2 \ = \ \half
\, , \label{O6condition}
\end{align}
where $\theta$ is an arbitrary phase parameter. 

The scalar components of vector multiplets and hypermultiplets
in ${\cal N}=2$ supergravity are governed by 
the special K\"{a}hler geometry and the quaternionic geometry,
respectively \cite{Andrianopoli:1996cm}. 
In type IIA theory compactified on Calabi-Yau three-fold,
the former (latter) geometry is described by the moduli space of the
K\"{a}hler form (the complex structure). 
In the theory compactified on generalized geometry,
such two geometries are given by the spaces 
${\cal M}_{\pm}$ discussed in the previous appendices
\cite{Benmachiche:2006df, Cassani:2007pq}.
Let us specify the supersymmetry truncation from 
${\cal N}=2$ to ${\cal N}=1$ via the O6 orientifold projection 
on generalized geometries \cite{Cassani:2007pq}.
To preserve half of the supersymmetry, we set $a = \ol{b} \, \e^{i \theta}$
as in (\ref{O6condition})
and project out some physical degrees of freedom:
\bsubeq \label{IIAtruncate}
\begin{align}
\xi^{\hat{I}} \ &= \ 0 
\ = \ {\rm Im} ({\cal C} Z^{\hat{I}}) 
\ = \ {\rm Re} ({\cal C} {\cal G}_{\hat{I}}) 
\, , \ls
\wt{\xi}_{\check{I}}
\ = \ 0 
\ = \ {\rm Re} ({\cal C} Z^{\check{I}}) 
\ = \ {\rm Im} ({\cal C} {\cal G}_{\check{I}}) 
\, , \label{IIAtruncate_hyper}
\end{align}
where the indices $I = 0,1,\dots, b^-$ are split into 
$I = (\hat{I}, \check{I})$. 
Due to this, each ${\cal N}=2$
hypermultiplet in type IIA is decomposed into two ${\cal N}=1$ chiral
multiplets with opposite spins.
In addition, ${\cal N}=2$ vector multiplets with 
indices $A = (0, a)$ are truncated as
\begin{gather}
A_{\mu}^{\check{A}} \ = \ 0
\, , \ls
X^{\hat{A}} \ = \ 0
\, , \label{IIAtruncate_vector} 
\\
{\cal F}_{\hat{A}} \ = \ 0 
\, , \ \ 
{\cal N}_{\check{A} \hat{B}} \ = \ 0
\, ; \ls
(K_+)_{\check{a} \ol{\hat{b}}} \ = \ 0
\, , \ \
D_{\check{a}} X^{\hat{B}} \ = \ 
D_{\hat{a}} X^{\check{B}} \ = \ 0
\, .
\end{gather}
\esubeq
Note that one has to truncate out the graviphoton $A^0_{\mu}$.
We split the indices $A = 0,1,\dots, n_v$ (where $n_v = b^+$) as
$\check{A} = 0,1,\dots, n_{ch}$ and $\hat{A} = 1,\dots,
\hat{n}_v = n_v - n_{ch}$ (with a restriction $n_v \geq n_{ch}$).
This means that ${\cal N}=2$ vector multiplets are decomposed into
${\cal N}=1$ vector multiplets and chiral multiplets 
with respective numbers ${n}_v$:
Some degrees of freedom are projected out in such a way as
$n_v \to \hat{n}_v$ in the vector multiplets, 
and as $n_v \to n_{ch}$ in the chiral multiplets.
Imposing (\ref{IIAtruncate_hyper}) and (\ref{IIAtruncate_vector}) 
on $\Phi_+$ and on $\Pi_-$, we obtain
$\Phi_+ = X^{\check{A}} \omega_{\check{A}} 
- {\cal F}_{\check{A}} \wt{\omega}{}^{\check{A}}$ 
and
$\Pi_- = U^{\check{I}} \alpha_{\check{I}} - \wt{U}_{\hat{I}} \beta^{\hat{I}}$, 
respectively.
Substituting them into the previous results, we write down the reduced functions:
\bsubeq \label{app-IIAKWN=1}
\begin{align}
{\cal W} \ &= \ 
- \frac{i}{4 \ol{a}b} \Big[
X^{\check{A}} \big( 
e_{\text{RR} \check{A}} - U^{\check{I}} e_{\check{I} \check{A}} 
+ \wt{U}_{\hat{I}} m_{\check{A}}{}^{\hat{I}} 
\big)
- {\cal F}_{\check{A}} \big( 
m_{\text{RR}}^{\check{A}} + U^{\check{I}} p_{\check{I}}{}^{\check{A}} 
- \wt{U}_{\hat{I}} q^{\hat{I} \check{A}} 
\big)
\Big]
\, , \label{app-IIASuperN=1} \\
K \ &= \ 
K_+ + 4 \varphi
\, , \label{app-IIAKahlerN=1} \\
K_+ \ &= \ 
- \log i \big( 
\ol{X}{}^{\check{A}} {\cal F}_{\check{A}} - X^{\check{A}} \ol{\cal F}_{\check{A}} 
\big) 
\, , \\
\e^{- 2 \varphi} \ &= \ 
\half \Big[ {\rm Im} ({\cal C} Z^{\check{I}}) {\rm Re} ({\cal C} {\cal G}_{\check{I}})
- {\rm Re} ({\cal C} Z^{\hat{I}}) {\rm Im} ({\cal C} {\cal G}_{\hat{I}}) \Big]
\, . \label{app-IIAdilatonN=1}
\end{align}
Substituting the truncation rules (\ref{IIAtruncate_hyper}) 
and (\ref{IIAtruncate_vector}) into (\ref{IIAD0}) with
setting $D^A \to D^{\hat{A}}$, we also obtain the D-term in ${\cal N}=1$ theory as
\begin{align}
D^{\hat{A}}
\ &= \ 
\e^{2 \varphi} [ ({\rm Im} {\cal N})^{-1} ]^{\hat{A} \hat{B}} 
\Big\{ {\rm Re} ({\cal C} Z^{\hat{I}}) 
\big[ e_{\hat{I}\hat{B}} + {\cal N}_{\hat{B}\hat{C}} p_{\hat{I}}{}^{\hat{C}} \big]
- {\rm Re} ({\cal C} {\cal G}_{\check{I}} ) 
\big[ m_{\hat{B}}{}^{\check{I}} + {\cal N}_{\hat{B}\hat{C}} q^{\check{I}\hat{C}} \big]
\Big\}
\, . \label{app-IIAD1}
\end{align}
\esubeq
We should notice that the D-term (\ref{app-IIAD1}) is a complex because
of the existence of an (anti-)holomorphic function ${\cal N}_{\hat{B} \hat{C}}$.
This appears in \cite{Andrianopoli:1996cm, Taylor:1999ii}.
This situation generically occurs when complex forms of flux
variables are turned on.
Then we should carefully define the scalar potential from this D-term.
We also substituted 
$c_- = |a|^2 - |b|^2 = 0$ by the O6 orientifold projection 
\cite{Cassani:2007pq}.

\end{appendix}


}
\end{document}